\documentclass{aa}  

\usepackage{graphicx}
\usepackage{txfonts}

\newcommand{\zt}{$\zeta$\,Tau}
\newcommand{\gc}{$\gamma$\,Cas}
\newcommand{\xmm}{\emph{XMM-Newton}}
\newcommand{\kms}{km\,s$^{-1}$}
\newcommand{\ch}{\emph{Chandra}}
\newcommand{\te}{TESS}

\begin{document}

   \title{X-raying the \zt\ binary system}

   \author{Ya\"el Naz\'e
          \inst{1}\fnmsep\thanks{F.R.S.-FNRS Senior Research Associate}
          \and
          Christian Motch\inst{2}
          \and
          Gregor Rauw\inst{1}
          \and
          Myron A. Smith\inst{3}
          \and
          Jan Robrade\inst{4}
          }

   \institute{Groupe d'Astrophysique des Hautes Energies, STAR, Universit\'e de Li\`ege, Quartier Agora (B5c, Institut d'Astrophysique et de G\'eophysique), \\
All\'ee du 6 Ao\^ut 19c, B-4000 Sart Tilman, Li\`ege, Belgium\\
              \email{ynaze@uliege.be}
         \and
             Universit\'e de Strasbourg, CNRS, Observatoire Astronomique de Strasbourg, 11 rue de l'Universit\'e, UMR 7550, F-67000 Strasbourg, France
         \and
         NSF OIR Lab, 950 N Cherry Ave, Tucson, AZ 85721, USA
         \and
         Hamburger Sternwarte, University of Hamburg, Gojenbergsweg 112, D-21029 Hamburg, Germany
            }


 
  \abstract
   {The Be star \zt\ was recently reported to be a \gc\ analog; that is, it displays an atypical (bright and hard) X-ray emission. The origin of these X-rays remains debated.}
   {The first X-ray observations indicated a very large absorption of the hot plasma component ($N_{\rm H}\sim10^{23}$\,cm$^{-2}$). This is most probably related to the edge-on configuration of the \zt\ disk. If the X-ray emission arises close to the companion, an orbital modulation of the absorption could be detected as the disk comes in and out of the line of sight.}
   {New \xmm\ data were obtained to characterize the high-energy properties of \zt\ in more detail. They are complemented by previous \ch\ and SRG/eROSITA observations as well as by optical spectroscopy and \te\ photometry.}
   {The high-quality \xmm\ data reveal the presence of a faint soft X-ray emission, which appears in line with that recorded for non-\gc\ Be stars. In addition, \zt\ exhibits significant short-term variability at all energies, with larger amplitudes at lower frequencies (``red noise''), as is found in X-ray data of other \gc\ stars. Transient variability (softness dip, low-frequency signal) may also be detected at some epochs. In addition, between X-ray exposures, large variations in the spectra are detected in the 1.5--4.\,keV energy band. They are due to large changes in absorption toward the hottest (9\,keV) plasma. These changes are not correlated with either the orbital phase or the depth of the shell absorption of the H$\alpha$ line. These observed properties are examined in the light of proposed \gc\ models.}
   {}


   \maketitle
%

\section{Introduction}
A prominent star in the night sky, \zt\ (B1Ve, $V$=3.03) has been known for more than a century to be a binary system with a $P\sim$133\,d orbit (see \citealt{rud09} and references therein). Its companion is a low-mass ($\sim$1\,M$_{\odot}$) object of unknown nature that could be a hot stripped Helium star or a white dwarf (WD). Such low-mass companions appear common for Be stars (e.g. \citealt{wan21,naz22bin}). In fact, in the current paradigm, they should naturally result from binary interactions that led to the formation of the fast-rotating Be star \citep{pol91,van97,sha14}.

The ``shell'' profile of the H$\alpha$ emission line of \zt, with its prominent central absorption, suggests a high inclination for its circumstellar disk. Values of 70--90$^{\circ}$ have been derived using H$\alpha$ fitting, polarimetric analyses, and/or interferometric measurements \citep{qui97,tyc04,gru06,gie07,car09,tou13,coc19}. As is common for Be stars, the usual disk tracer, the H$\alpha$ profile, evolves over decades, with minimal emissions being notably recorded at the end of the 1980s and around 2013 \citep{rud09,pol17}. In addition, changes in the violet-to-red (V/R) peak ratio and/or the central absorption strength have also been detected \citep{rud09,ste09,naz22}. A complicated profile, with multiple emission peaks, is also sometimes observed \citep{rud09,ste09}. It can be noted that such profiles can be interpreted as either triple emissions or a double absorption \citep{ste09}. In the 1990s and 2000s, the H$\alpha$ line of \zt\ presented a stable cycle of about 1400\,d. It was interpreted as the effect of a precessing disk \citep{sch10} or of a one-arm spiral oscillation in a viscous disk \citep{car09}. More recently, the variations changed, with little V/R variability but a strong modulation of the central absorption depth \citep{naz22}. Contrary to older observations, no periodicity was detected; rather, there appears to be some cyclicity over timescales of hundreds of days.

The X-ray emission of \zt\ was examined only recently. Our study of Be+sdOB binaries or candidates used an archival \ch\ 10\,ks-exposure (ObsID=26239), in which \zt\ appeared heavily piled-up\footnote{``Pile-up'' in X-ray observations occurs when several photons impact the same pixel or neighbouring pixels during one observing frame (i.e. one read-out cycle). These photons are then erroneously recorded as a single photon with an energy corresponding to the sum of those of the incoming photons. One notable consequence is a spectral distortion, with the X-ray source appearing harder than it really is. } \citep{naz22}. To overcome this problem, the data were extracted in an annulus rather than a circle. This degraded the signal-to-noise but allowed us to examine the X-ray emission, for which the presence of a strongly absorbed hot X-ray plasma was clearly detected. The high luminosity combined with the high temperature of the X-ray emitting plasma revealed that \zt\ belongs to the growing class of \gc\ analogs. This \gc\ character was subsequently confirmed thanks to the first four semesters of eROSITA observations \citep{naz23}, although the short exposure time (even when combining all semesters) and the presence of optical loading (i.e., contamination of X-ray data by optical and/or UV photons) limited the quality of these data.

\gc\ analogs share unique X-ray characteristics \citep{smi16,naz18}: high temperatures ($kT>5$\,keV), high luminosities ($\log(L_{\rm BOL}/L_{\odot})$ of --6.2 to --4, which are in between values for ``normal'' OB stars and X-ray binaries), short-term ``shots'', and the presence of an iron fluorescence line at 6.4\,keV. To date, peculiarities have only been found for these stars in the X-ray range. The origin of the \gc\ phenomenon remains a puzzle. Several scenarios have been proposed to explain the presence of such a hot plasma. One of them envisages the X-rays arising from localized magnetic star-disk interactions \citep{smi98}. Others rather invoke a major role for the companion to the Be star, with either accretion onto a WD \citep{ham16}, accretion onto a neutron star in the propeller stage \citep{pos17}, or a collision between the wind of a subdwarf companion and the Be disk \citep{lan20}. Both observational and theoretical arguments could discard the latter two ideas \citep{smi17,naz22,rau24}. However, the non-detection of subdwarf stars in some \gc\ systems, suggesting that the companions are WDs, may provide some support for the accreting WD idea \citep{gie23}. Two scenarios thus remain, with or without the (contemporary) involvement of a companion.

To better characterize the high-energy emission of \zt, and gain further insight into the origin of the X-ray emission, we obtained two high-quality \xmm\ exposures of \zt. They were taken at two very different orbital phases, allowing us to probe the system first when the Be star appears in front of its companion and then in the opposite situation. If the X-rays arise close to the companion and the absorption is due to the Be disk, then a strong modulation of the absorption with orbital phase should be detected - but alternative configurations are possible. The \xmm\ data will therefore be interpreted considering the previous X-ray observations but also optical spectroscopy and broad-band photometry from \te. The data are presented in Section 2, the observational results are provided in Section 3 and discussed in Section 4, and Section 5 summarizes our conclusions.

\section{Data}

\begin{figure}
  \centering
  \includegraphics[width=9cm]{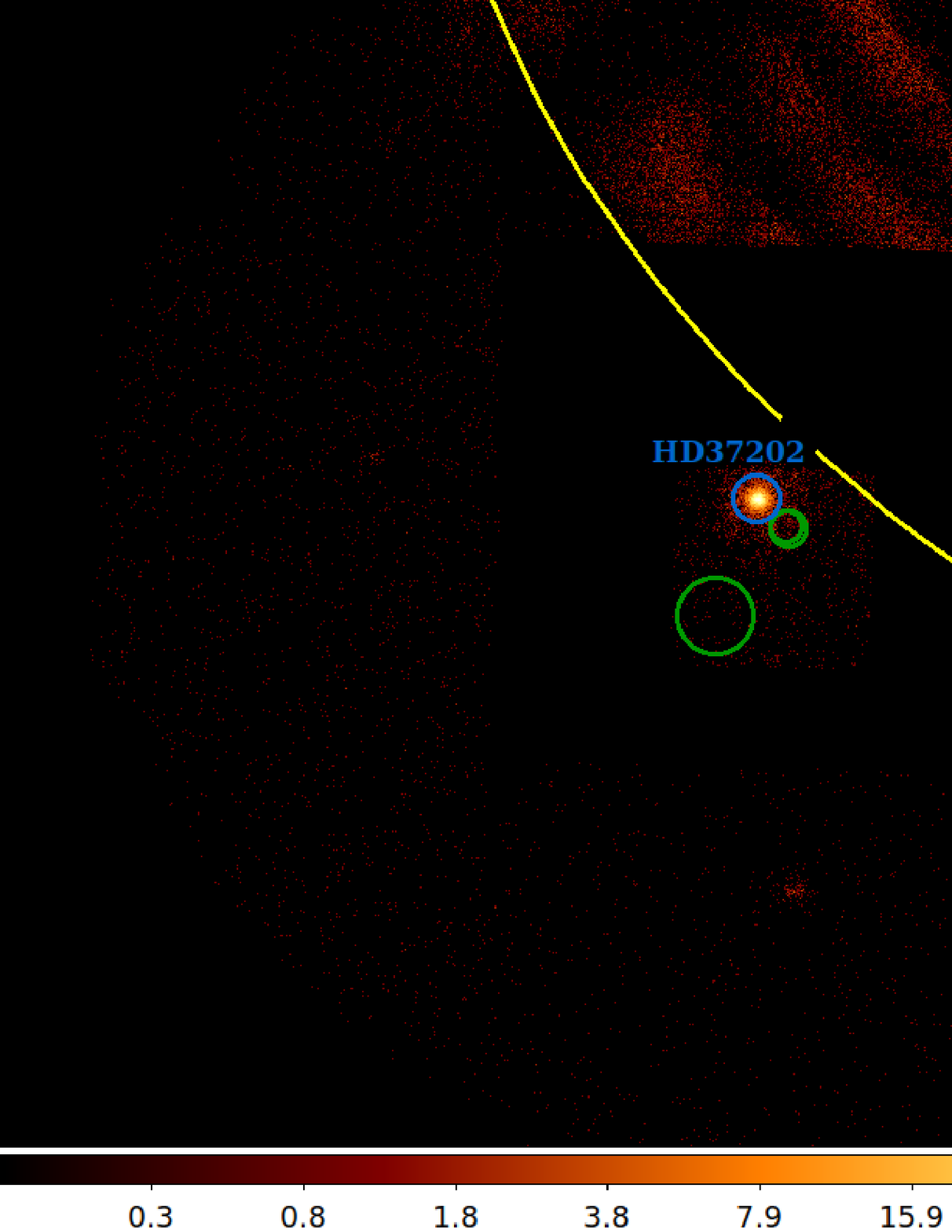}
  \includegraphics[width=9cm]{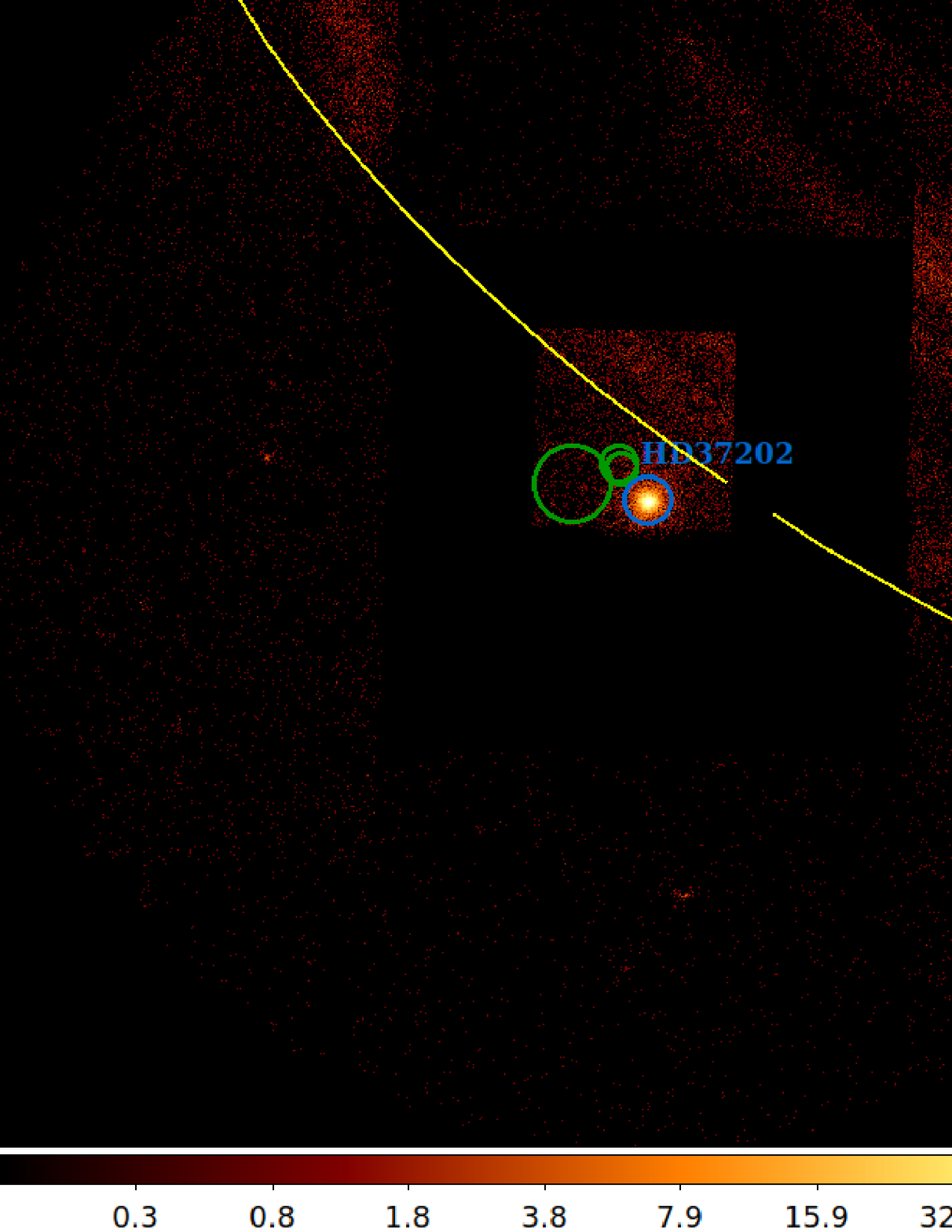}
  \caption{Field of view (15\arcmin\ radius) of both \xmm\ exposures (all EPIC cameras combined; March 2023 on top, October 2023 at the bottom). The yellow arcs mark the limit of the stray light contamination, the blue circles correspond to the source extraction regions (30\arcsec\ radius), and the green circles are the background extraction regions (their size is limited by the ``small window'' mode: the smaller ones are for MOS cameras, the larger one for the pn camera).}
  \label{fov}
\end{figure}
   
The \xmm\ data were acquired in March and October 2023 (ObsID=0920020301, 0920020401). These epochs correspond to two opposite orbital phases, $\phi=0.0$ and 0.5 (Be star in front and companion in front, respectively, using the ephemeris of \citealt{naz22}). The observations used a thick filter to avoid optical contamination from the very bright optical emission of that object ($V=3.03$). In addition, the ``small window'' mode was used to avoid pile-up.

The data reduction was done with the Science Analysis Software (SAS) v21.0.0 and calibration files available in September 2023. First, the pipeline processing was applied to both European Photon Imaging Camera (EPIC) cameras, then a filtering was done on the event files to keep only the best-quality data ({\sc{pattern}} 0--12 for MOS and 0--4 for pn). Following the inspection of global light curves at energies above 10\,keV, no contamination by background proton flares was detected.

\begin{figure*}
  \centering
  \includegraphics[width=6cm]{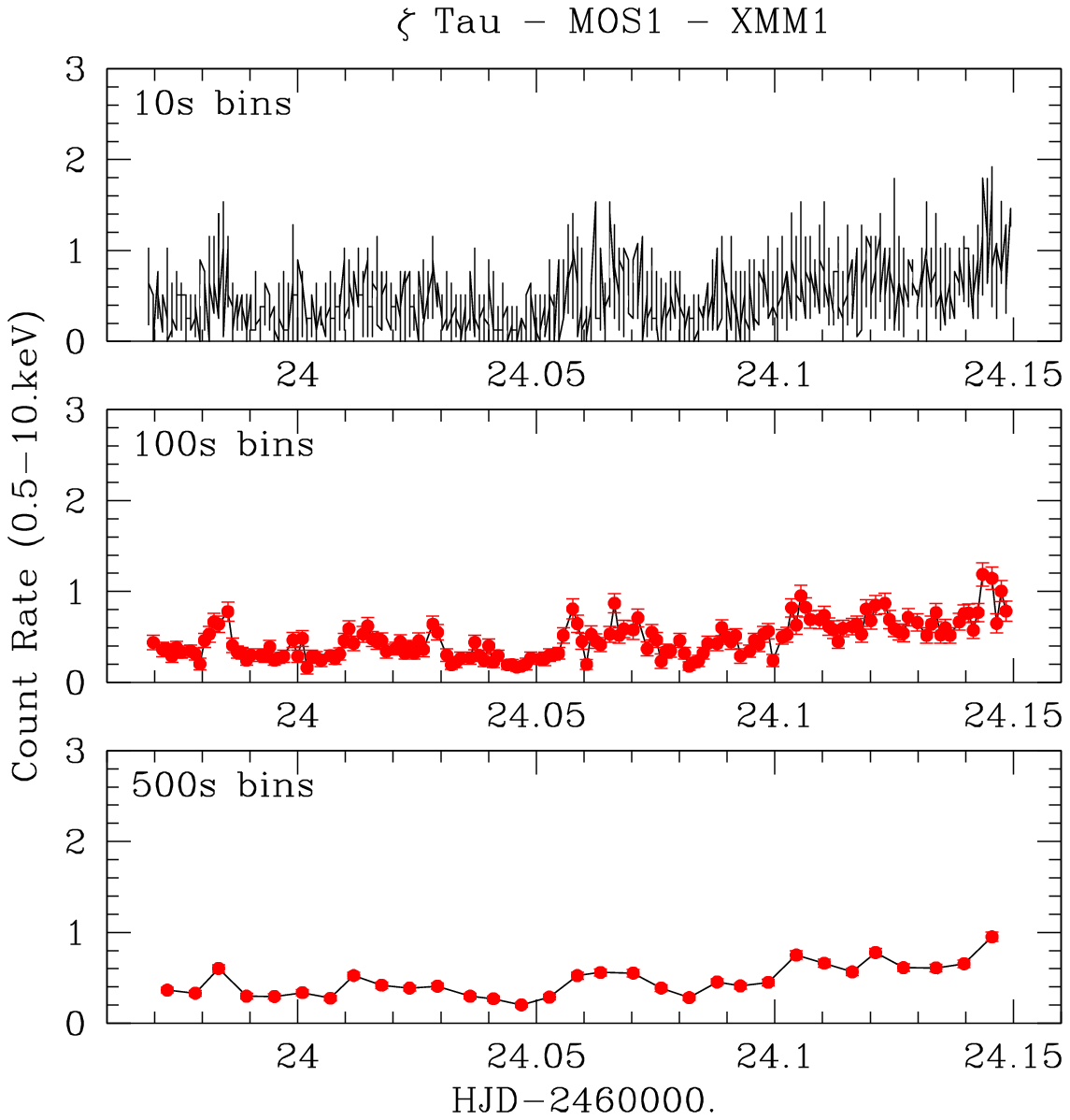}
  \includegraphics[width=6cm]{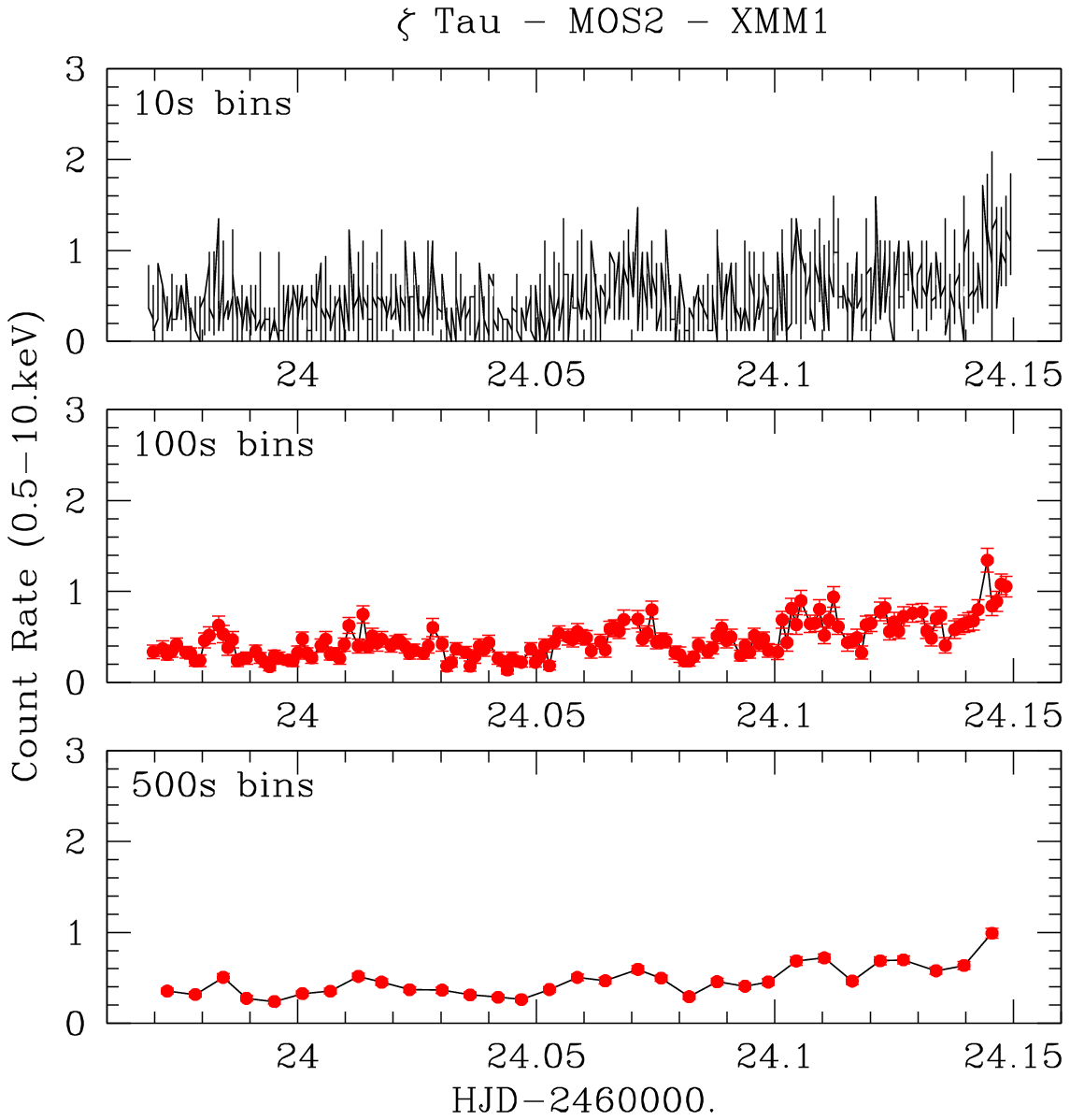}
  \includegraphics[width=6cm]{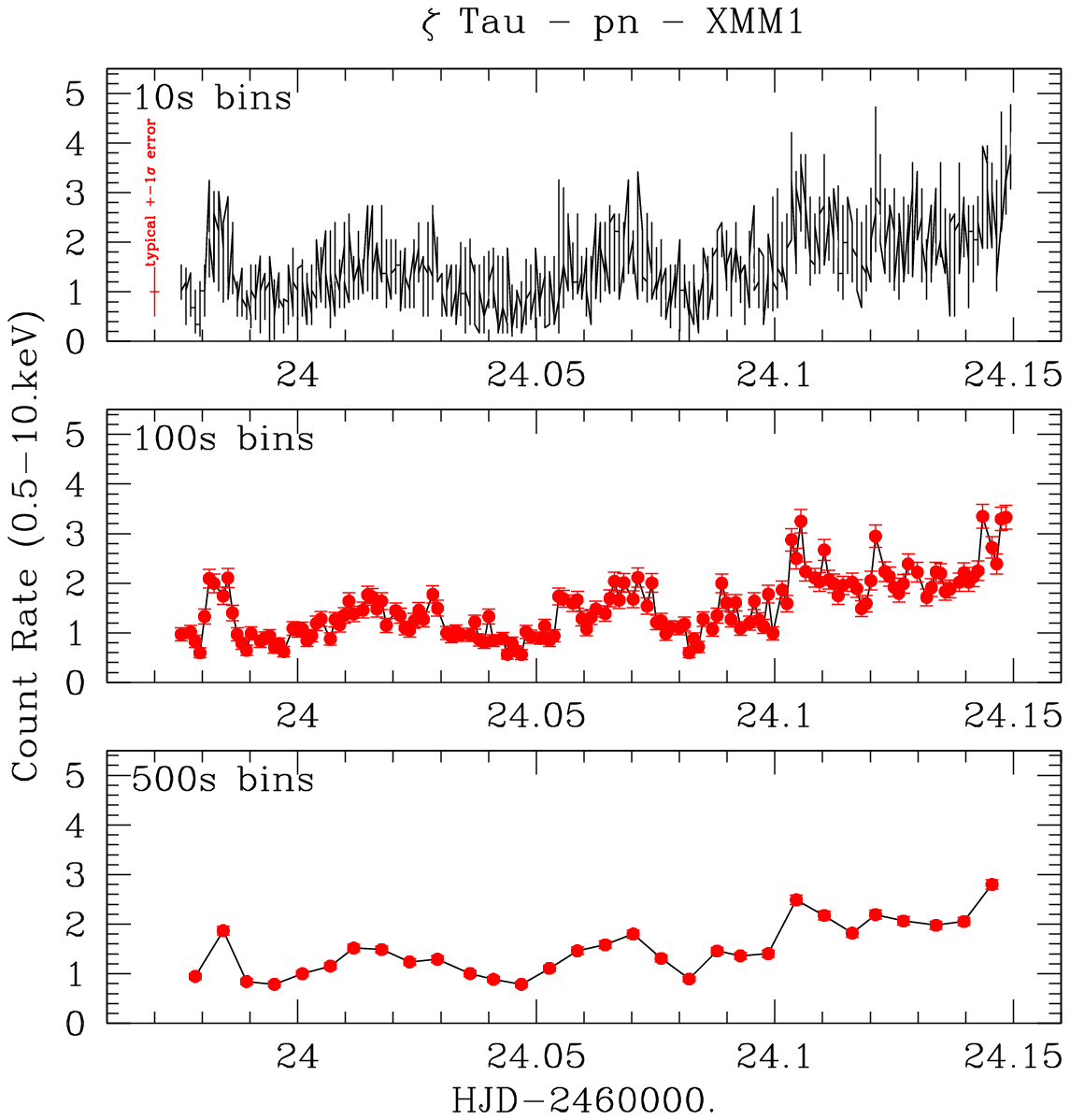}
  \includegraphics[width=6cm]{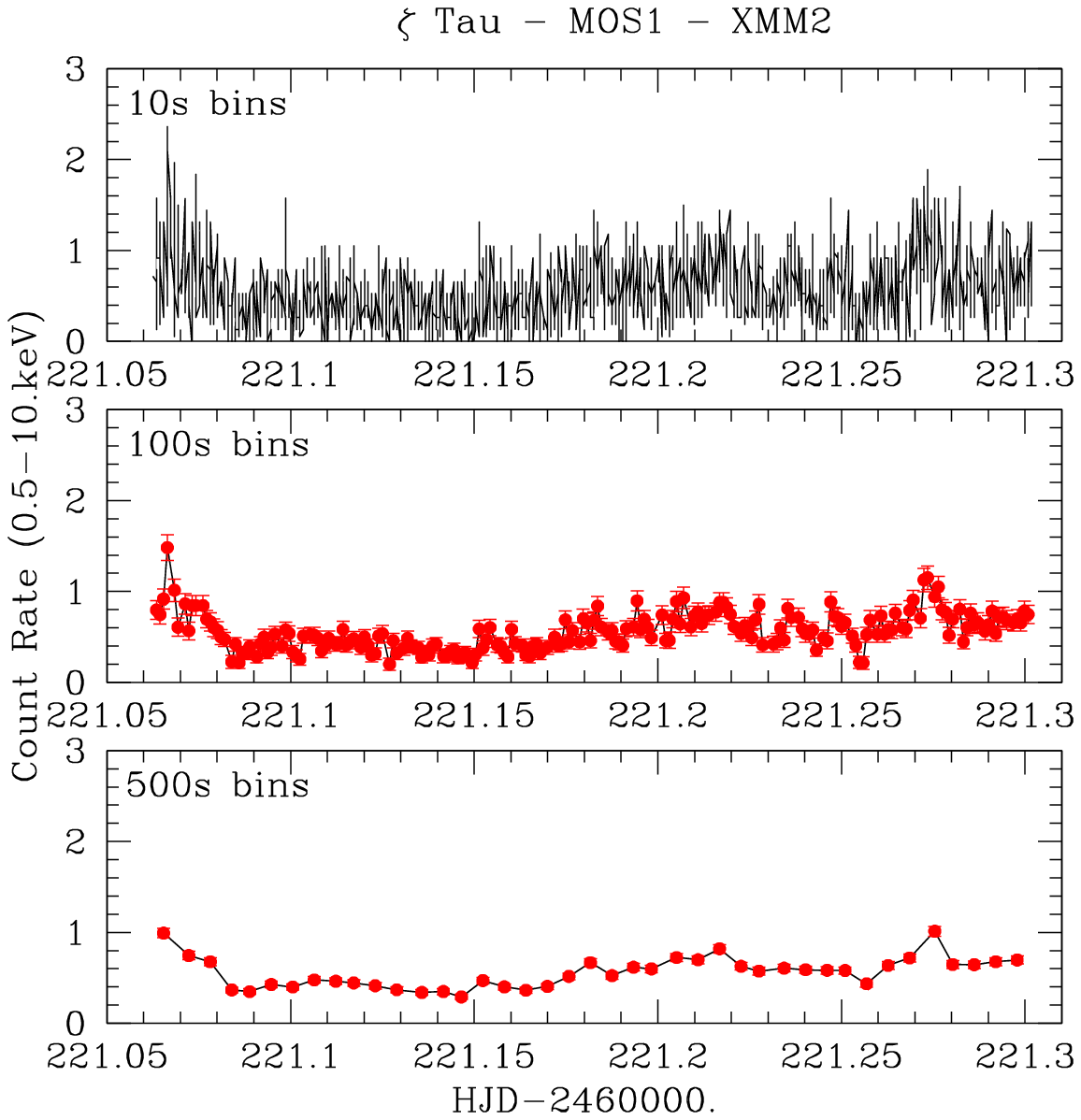}
  \includegraphics[width=6cm]{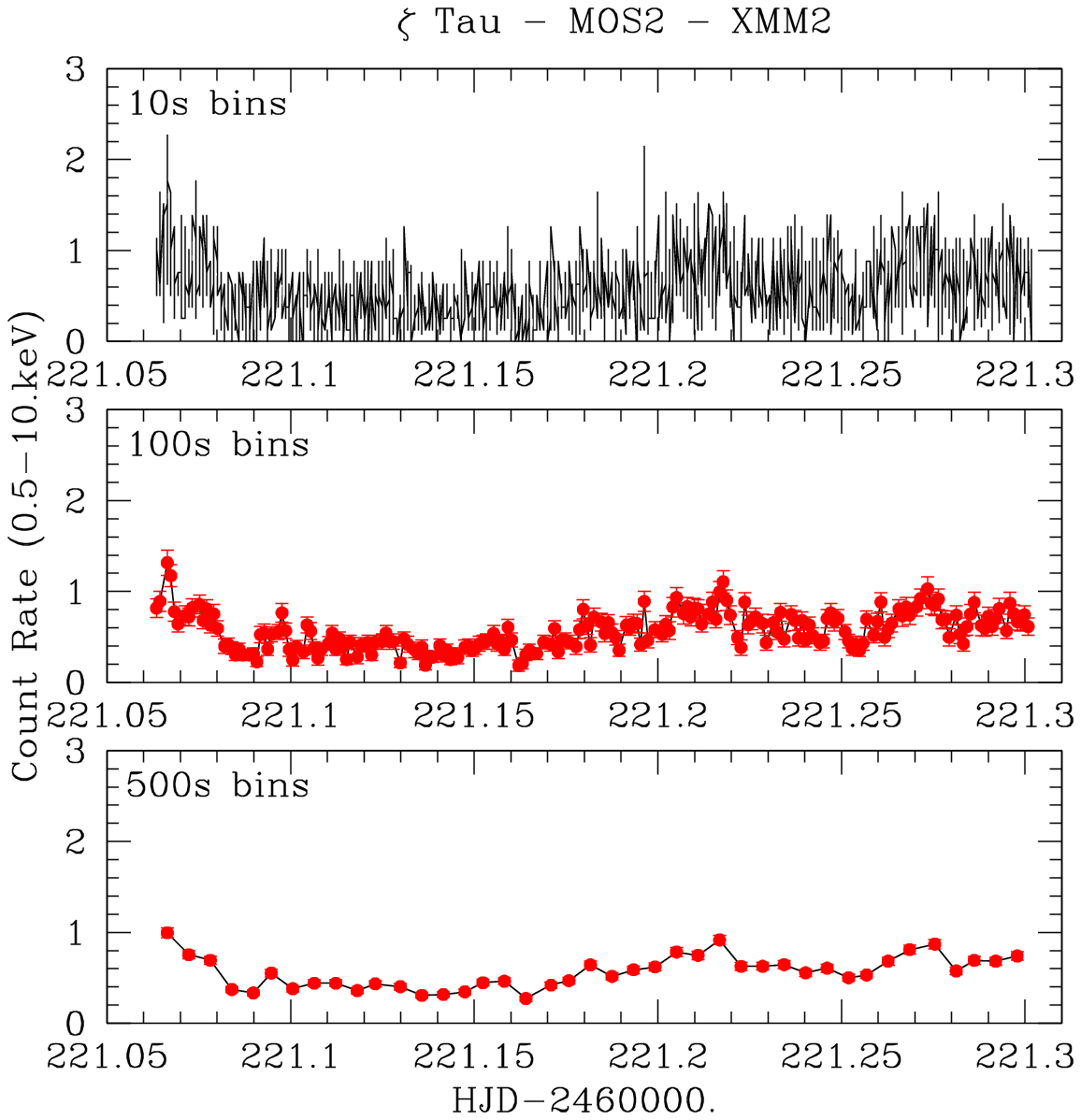}
  \includegraphics[width=6cm]{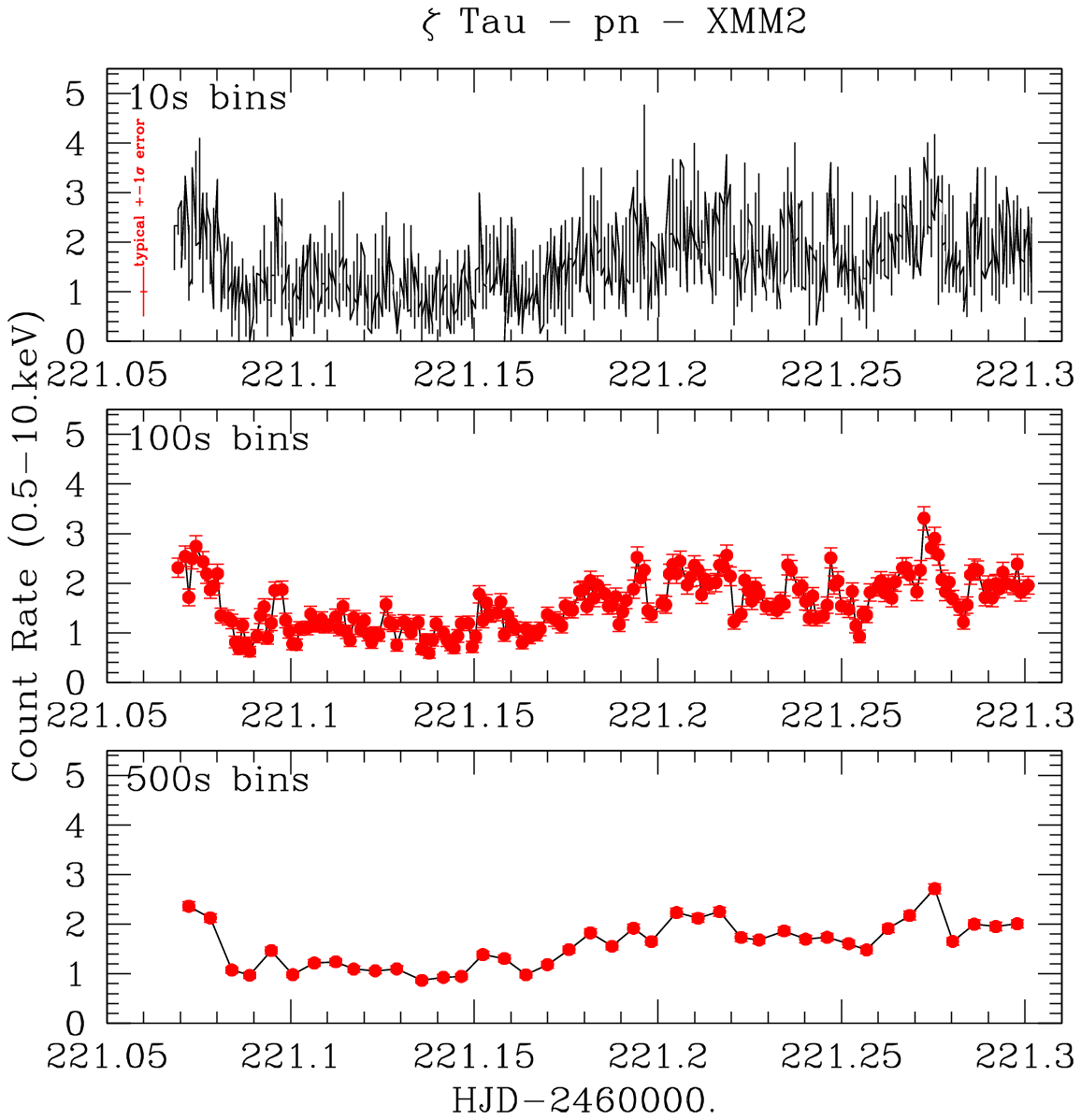}
  \caption{X-ray light curves recorded for \zt\ with EPIC-MOS1 (left), MOS2 (middle) and pn (right) in the 0.5--10.\,keV energy band and for three time bins. The data points and their error bars are shown in red for the two longer bins. The first \xmm\ observation is shown on top and the second one at bottom. 
  }
  \label{lc}
\end{figure*}
   
Stray light from the Crab Nebula, which lies at 1.1$^{\circ}$ of \zt, contaminates the field of view (Fig. \ref{fov}). Fortunately, it does not impact \zt\ or its close surroundings so that a clean extraction can be done for the target. The source counts were extracted in a circle centered on the Simbad coordinates of \zt\ and with a radius of 30\arcsec. Background regions were chosen on the same central CCD, at a location as nearby as possible from the source and with the largest radius possible (see Fig. \ref{fov}).

Using these regions and the dedicated SAS reduction tasks, spectra and their response matrices were built, as well as light curves corrected to provide equivalent full-PSF, on-axis count rates. To assess the effect of time binning, light curves were calculated for 10, 100, and 500\,s bins in the 0.5--10.\,keV energy band. In addition, to investigate changes in hardness, light curves were constructed for two energy bands, 0.5--2.\,keV and 2.--10.\,keV with the 500\,s bin. To investigate the presence of ``softness dips'' \citep{ham16}, three additional energy bands were considered (0.5--1.5, 1.5--4., 4.--10.\,keV) but only with a 1000\,s time bin in view of the low number of counts in the soft band. These bands also correspond to three zones with different long-term behavior (see Sect. 3.2). The average count rates in the total band are 0.41 and 1.32\,cts\,s$^{-1}$ for MOS and pn in the first exposure, respectively, and 0.51 and 1.45\,cts\,s$^{-1}$ for the same cameras in the second exposure. The counts in 0.5--2.\,keV represented only about 2\% of all recorded counts in the first exposure, but that fraction doubled in the second one.

\setcounter{figure}{3}
\begin{figure*}
  \centering
  \includegraphics[width=6cm]{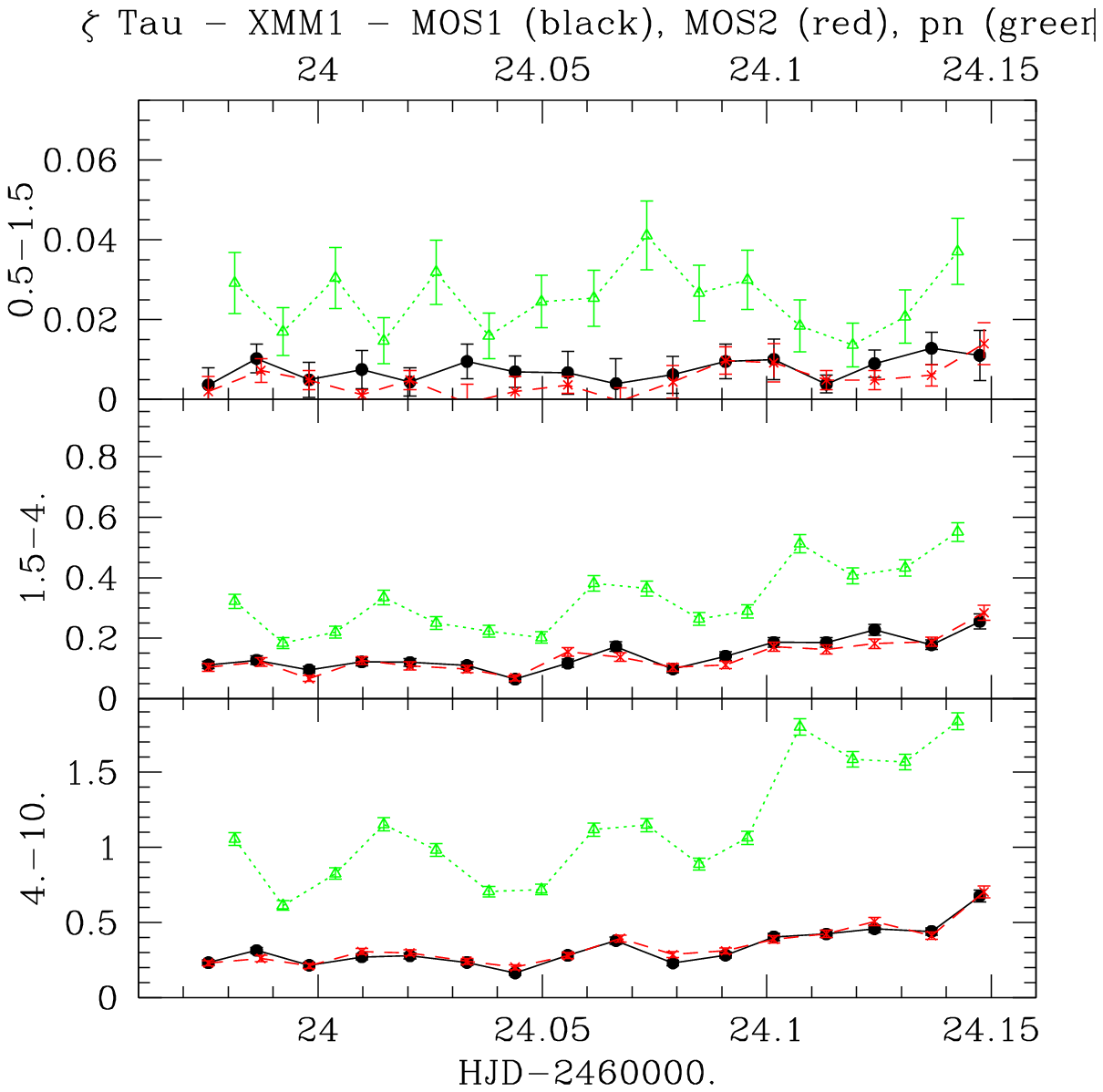}
  \includegraphics[width=6cm]{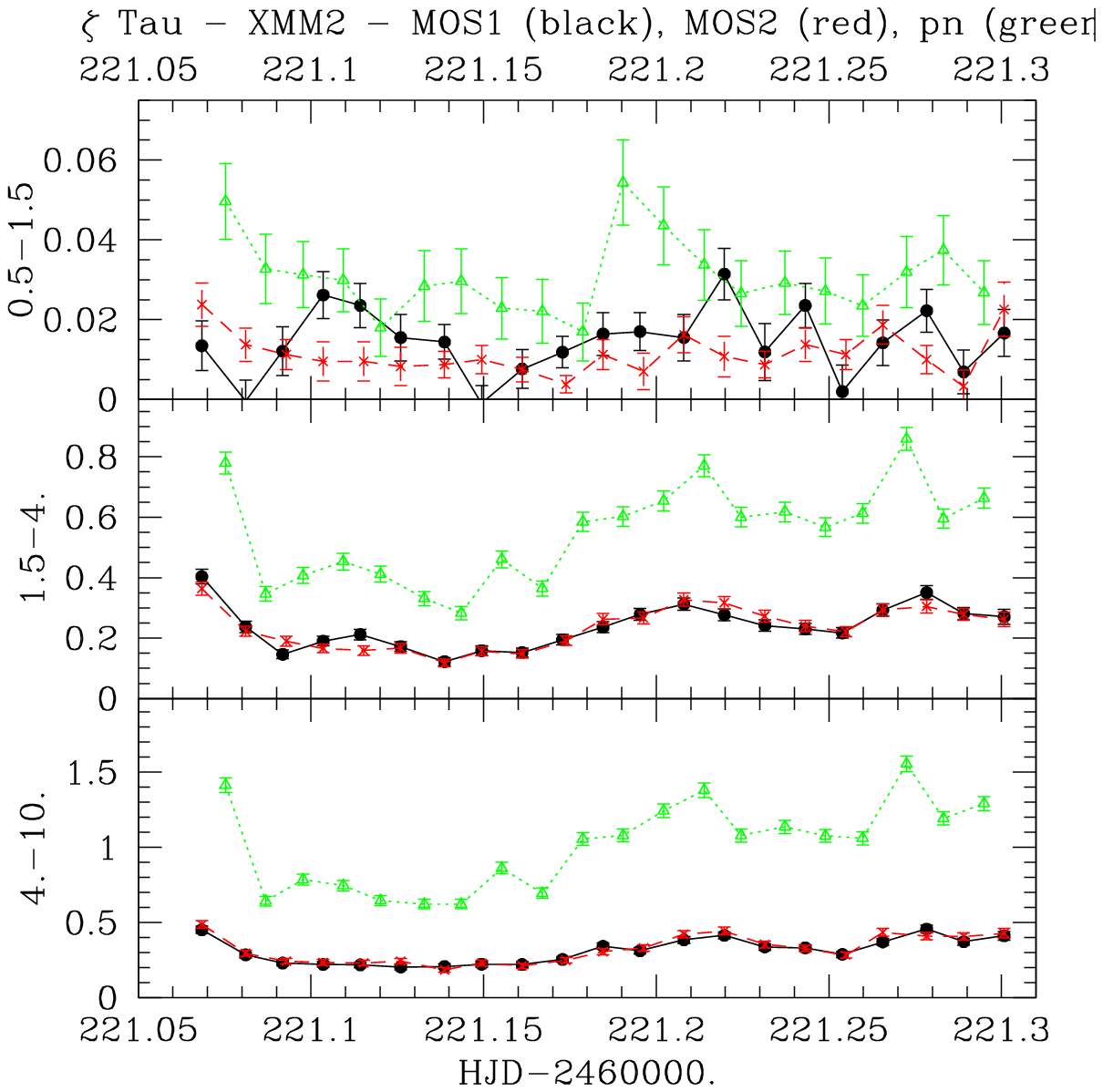}
  \includegraphics[width=6cm]{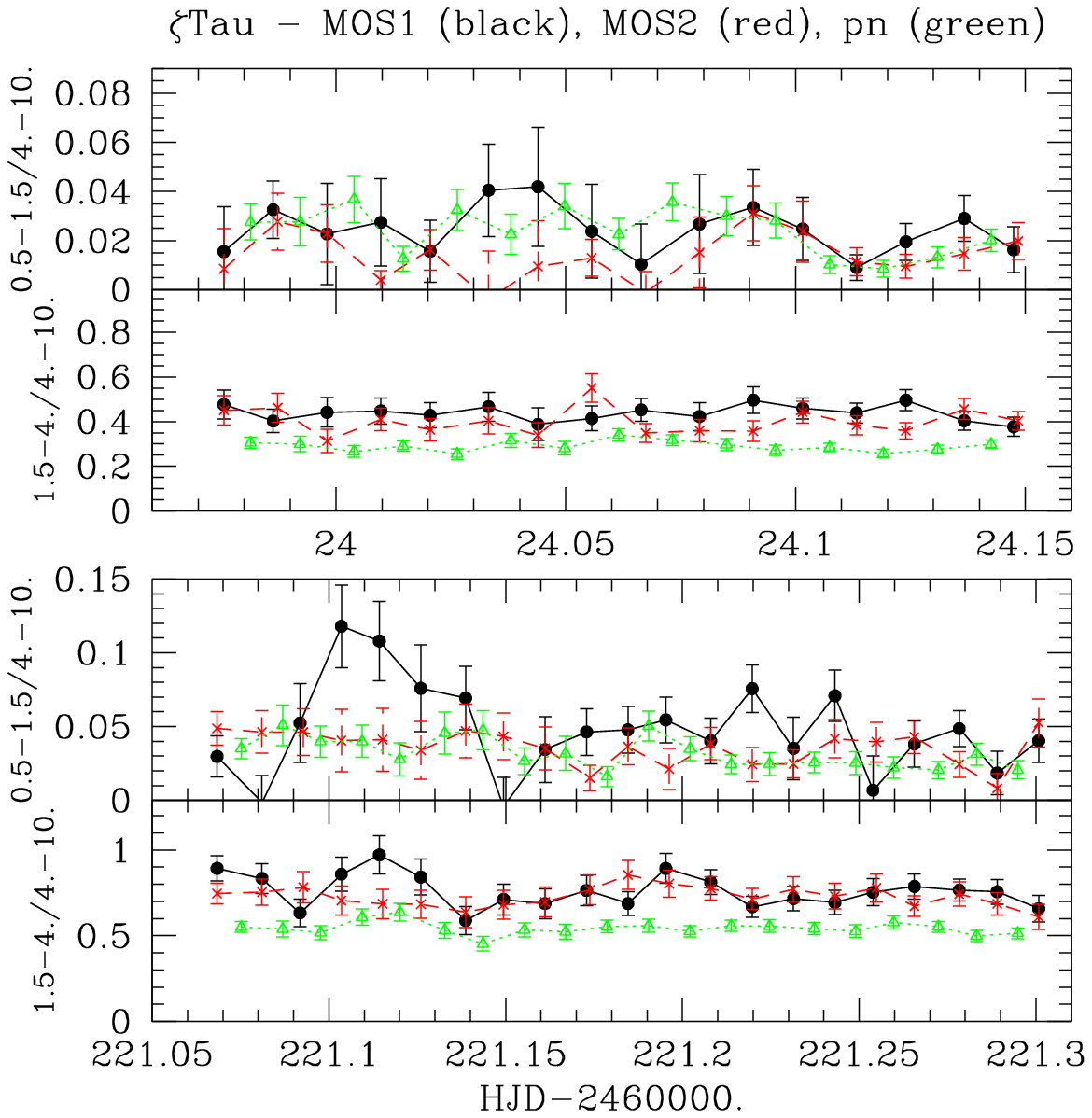}
  \caption{Curves in three energy bands and their ratios. {\it Left and Middle:} EPIC light curves in the two \xmm\ observations (March in the left panel, October in the middle panel) in the 0.5--1.5, 1.5--4., and 4.-10.\,keV energy bands with a 1\,ks time bin. EPIC-MOS1 data are shown with a solid black line and dots, MOS2 with a dashed red line and crosses, and pn with a dotted green line and triangles. {\it Right:} Ratios of these count rates for the first (top) and second (bottom) \xmm\ observations.}
  \label{lc2b}
\end{figure*}

In support of the \xmm\ data, we made use of the \ch\ and eROSITA data previously published \citep{naz22,naz23}. In addition, optical amateur data presented in \citet{naz24}, and taken close to the time of the X-ray exposures, are also used. Finally, \zt\ has been observed by the \te\ mission, with a 2\,minute cadence, during Sectors 43 to 45 (in 2021) as well as Sectors 71 and 72 (in 2023). These light curves were downloaded from MAST archives\footnote{https://mast.stsci.edu/portal/Mashup/Clients/Mast/Portal.html} and filtered to keep only the corrected points of highest quality (null quality flag and PDCSAP - Pre-search Data Conditioning Simple Aperture Photometry - to eliminate instrumental, long-term trends). Fluxes were converted into magnitudes and the average value was subtracted.

\section{Results}

\subsection{Intra-pointing light curves}
Figure \ref{lc} shows the X-ray light curves in the total energy band for the three EPIC cameras and both exposures. All three cameras display a similar behavior, with variations well over the expected Poisson noise. In the first \xmm\ observation, the count rate increases over the whole duration of the exposure. In addition, several short-term flux increases can be discerned: a narrow one at the start of the exposure, followed by approximately three broader ones. In the second observation, we can see the end of such an event at the start of the exposure, followed by a shallow general increase superimposed on several narrow increases (``shots'' and shot aggregates) of various amplitudes and durations. Such features are regularly seen in \gc\ analogs \citep{lop10,tor12,naz17} and can even be considered as typical of this class, as they are not detected in ``normal'' OB stars \citep{smi16}. 

\setcounter{figure}{2}
\begin{figure}
  \centering
  \includegraphics[width=9cm,bb=19 330 593 718, clip]{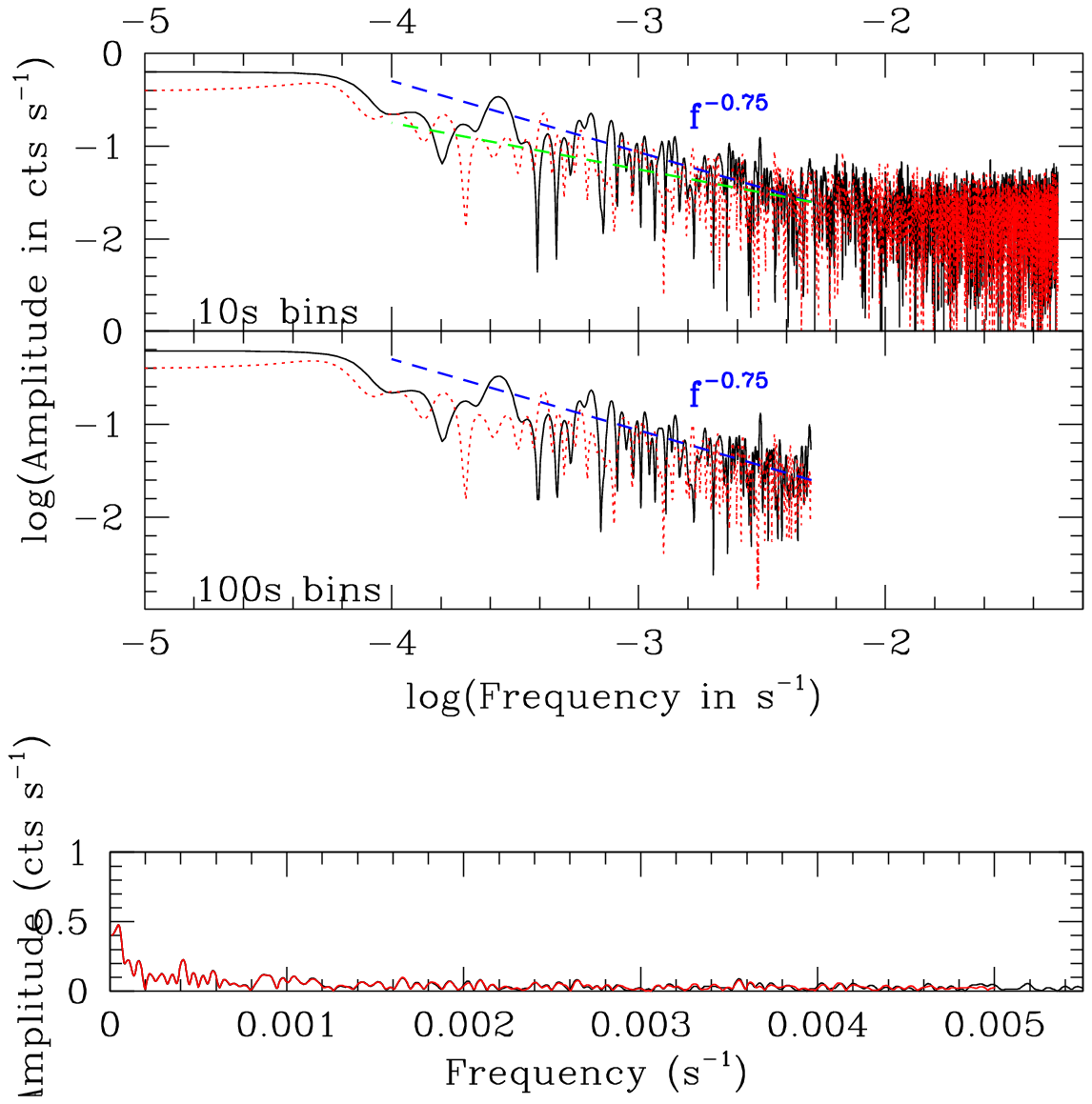}
  \caption{Fourier transform of the EPIC-pn light curves in total band, shown with log-log scaling. The 10 and 100\,s binning are used on top and bottom, respectively, and the first and second \xmm\ observations are represented by the solid black and dotted red lines, respectively. The blue line provides $A\propto f^{-0.75}$ while the green line shows $A\propto f^{-0.5}$. }
  \label{four}
\end{figure}
\setcounter{figure}{4}

A Fourier algorithm \citep{gos01} was applied to the total band curves for both 10 and 100\,s bins, which yields similar results (Fig. \ref{four}). No persistent periodic signal is detected although a small (transient) peak appears at a timescale of 3.7\,ks (about one hour) in the first \xmm\ observation. As in other \gc\ analogs \citep{lop10,tor12,naz17}, a decreasing amplitude as the frequency increases is detected. The trend follows $A\propto f^{-0.75}$, which is slightly steeper than the $A\propto f^{-0.5}$ found for \gc\ in \citet{lop10} and \citet{smi16}, but the periodogram is not as precise as in \gc\ so that the exponent is less well constrained and an exponent of --0.5 cannot be excluded. At higher frequencies, Poisson noise comes into play and the upper limit of the curve flattens, so that the two-slope break near 0.01\,Hz seen in \gc\ \citep{lop10} cannot be detected here.

Turning to the hardness ratios, no significant correlation between the count rate in the total band and hardness can be discerned. However, \citet{ham16}, \citet{smi19}, and \citet{rau22} reported on the presence of ``softness dips'' in the light curve of \gc; that is, time intervals with a lower fraction of very soft counts in the signal (whatever its overall amplitude). A feature of this type may have been possibly detected at the end of the first \xmm\ observation, although it was detected only in the pn camera and with a low significance (Fig. \ref{lc2b}). We note however that the soft count rate is very low and the noise in the MOS cameras is larger than for the pn.

\begin{figure}
  \centering
  \includegraphics[height=8.5cm]{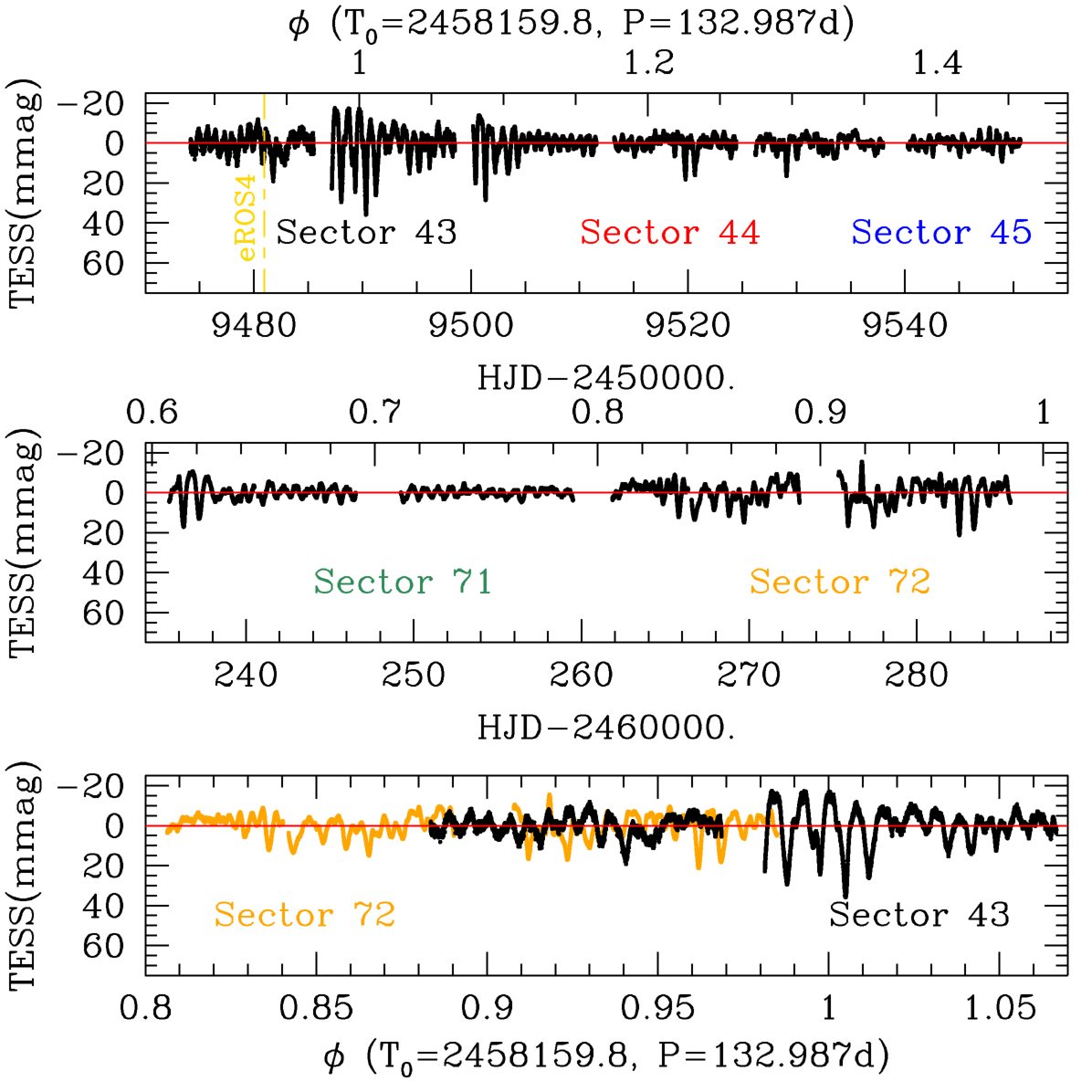}
  \includegraphics[height=8.5cm]{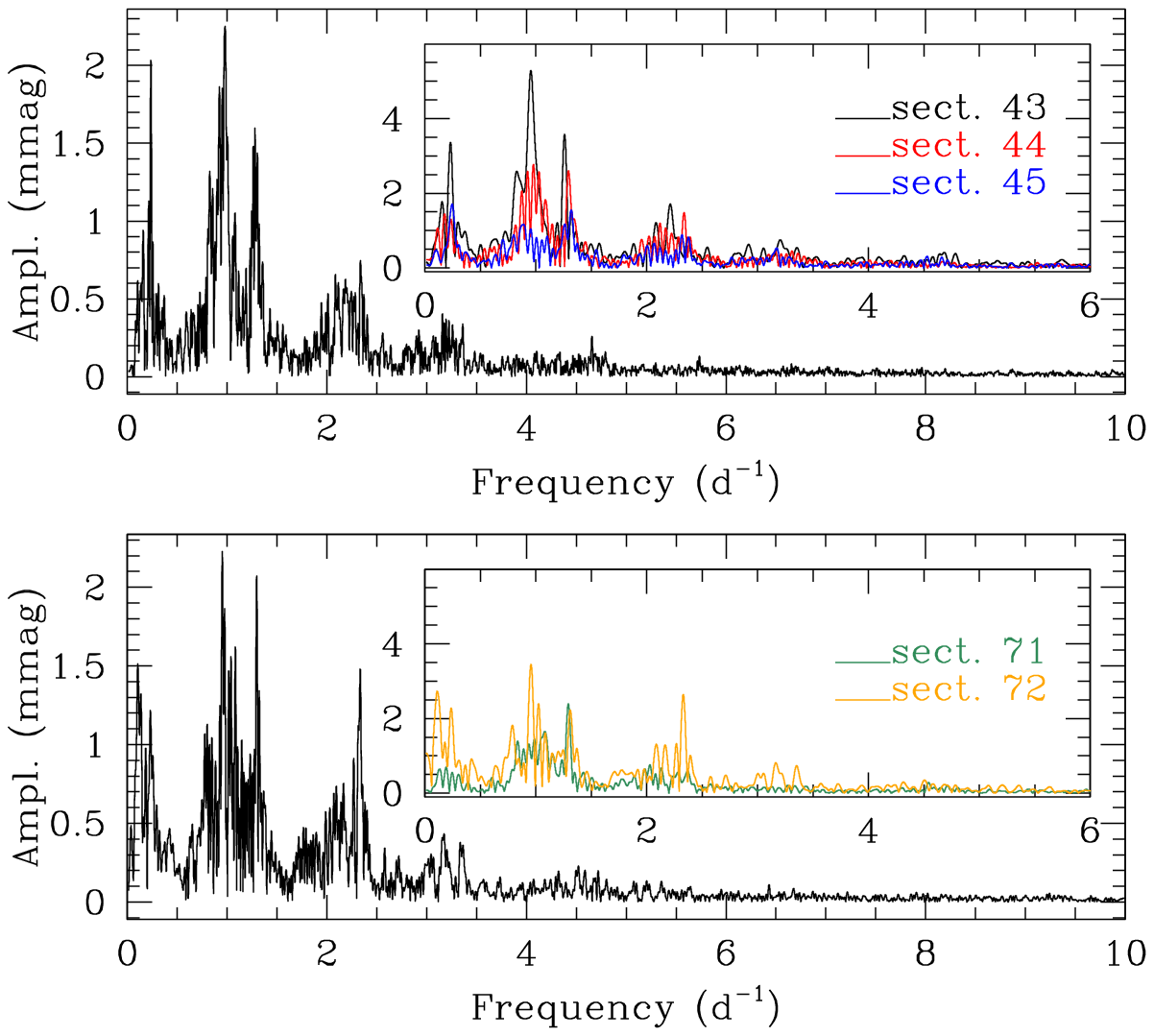}
  \caption{\te\ data of \zt. {\it Top:} The \te\ light curves of \zt\ in sectors 43--45 (first panel) and 71--72 (second panel), with phase axis on top and Julian Date axis at the bottom. The third panel compares the data overlapping in orbital phase. {\it Bottom:} Periodograms of the combined (43--45 and 71--72) light curves with insets showing the individual periodograms for each Sector.}
  \label{tess}
\end{figure}
   
\subsection{\te}

The \te\ light curves of \zt\ are shown on Fig. \ref{tess}. The amplitude of the variations is rather small, generally below 10\,mmag as in other \gc\ analogs \citep{naz20}. However, the light curve shape is not constant as there are short intervals at which oscillations clearly display larger amplitudes. The transition to such changes appears rather abrupt. The periodograms for individual as well as combined sectors has no significant high-frequency signal (10--360\,d$^{-1}$) but display broad multi-frequency peaks at low frequencies (``frequency groups'') as often found in Be stars \citep{naz20,lab22}. These periodograms display similar, but not exactly identical, features. In particular, the amplitude of the signals do considerably vary with the considered sector, as could be expected from the light curve behavior. The structure of the frequency groups also changes from sector to sector. The most common signals are detected near 0.23--0.24\,d$^{-1}$, near 0.95--1.08\,d$^{-1}$, near 1.26--1.32\,d$^{-1}$, and near 2.0--2.5\,d$^{-1}$.

All \te\ photometric datasets was gathered as the H$\alpha$ line strength was declining (i.e. its core absorption was growing, see Section 4.1), but there does not seem to be any simultaneous monotonic trend in the \te\ data. The fourth eROSITA scan was performed at the start of the 2021 \te\ observations, just before the strong oscillation episode. Nothing particular is detected in the optical light curve at the time of these X-ray observations (Fig \ref{tess}). The last \xmm\ observation was taken just before the 2023 \te\ data were gathered hence nothing can be concluded regarding the optical behavior at that specific epoch. 

The largest amplitude of the oscillations is detected near $\phi=0$. The 2021 and 2023 \te\ datasets cover different orbital phase intervals, but there is a small phase overlap near $\phi=0.9-0.95$. When those two datasets are directly compared (Fig \ref{tess}), the \te\ data show no sign of a recurrent behavior with orbital phase: the zero phase of the strongest oscillations in 2021 thus probably is a coincidence.

\begin{figure}
  \centering
  \includegraphics[height=8.5cm,angle=270]{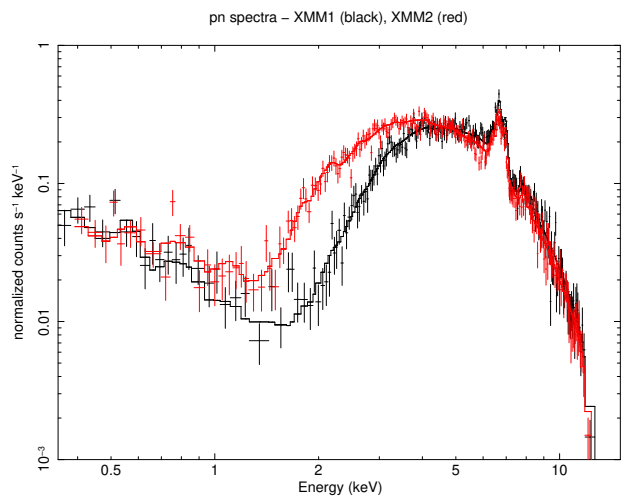}
  \caption{X-ray spectra recorded for \zt\ by the EPIC-pn camera in the first (black) and second (red) \xmm\ observations, along with their best-fit models (top of Table \ref{fit}). }
  \label{xmms}
\end{figure}
   
\begin{table*}
  \centering
  \caption{Results of the spectral fitting of \xmm\ spectra. }
      \label{fit}
 \begin{tabular}{lcccccccc}
 \hline\hline
 ObsID & $HJD-2.4e6$ & $\phi$ & $norm_1$ & $norm_2$ & $kT_3$ & $norm_3$ & $N_{\rm H}$ & $norm_4$ \\ 
 &             &        & ($10^{-4}$\,cm$^{-5}$) & ($10^{-6}$\,cm$^{-5}$) & (keV) & ($10^{-5}$\,cm$^{-5}$) & ($10^{22}$\,cm$^{-2}$) & ($10^{-2}$\,cm$^{-5}$) \\
2nd row&&& $norm_G$ & $\chi^2$/dof & \multicolumn{3}{c}{$F_{\rm X}^{obs}$ in erg\,cm$^{-2}$\,s$^{-1}$} \\
&&& ($10^{-5}$\,ph &               & (0.5--10.) & (0.5--2.) & (2.--10.) \\
&&& cm$^{-2}$\,s$^{-1}$) &    &  ($10^{-11}$) & ($10^{-14}$) & ($10^{-11}$)\\
 \hline
0920020301 &60024.054 &0.02 &3.55$\pm$0.40 & 6.05$\pm$1.51 & 4.16$\pm$1.80 & 2.95$\pm$3.03 & 17.21$\pm$0.20 & 2.63$\pm$0.03 \\ &&& 2.52$\pm$0.31 & 648.87/565 & 1.77$\pm$0.02 & 4.65$\pm$0.56 & 1.76$\pm$0.01 \\
0920020401 &60221.177 &0.50 &2.49$\pm$0.37 & 7.51$\pm$1.65 & 2.58$\pm$1.39 & 4.16$\pm$1.13 &  9.41$\pm$0.10 & 2.00$\pm$0.02 \\ &&& 2.60$\pm$0.25 & 854.63/658 & 1.70$\pm$0.01 & 11.5$\pm$0.7  & 1.69$\pm$0.01 \\
\hline
0920020301 &60024.054 &0.02 &3.56$\pm$0.40 & 6.10$\pm$1.50 & $=kT_4$ & 3.44$\pm$0.77 & 17.26$\pm$0.19 & 2.64$\pm$0.03 \\ &&& 2.51$\pm$0.31 & 649.12/566 & 1.77$\pm$0.03 & 4.68$\pm$0.26 & 1.76$\pm$0.01 \\
0920020401 &60221.177 &0.50 &2.43$\pm$0.36 & 7.05$\pm$1.62 & $=kT_4$ & 5.87$\pm$1.01 &  9.47$\pm$0.09 & 2.01$\pm$0.02 \\ &&& 2.59$\pm$0.25 & 858.86/659 & 1.70$\pm$0.01 & 11.6$\pm$0.4  & 1.69$\pm$0.01 \\
\hline
 \end{tabular}        
 \tablefoot{The absorbed thermal model has the form $phabs(ISM)\times[apec_1+apec_2+apec_3+phabs\times apec_4+Gaussian]$. The ephemerids of \citet{naz22} are used, i.e. $P=132.987$\,d and $T_0=2\,458\,159.8$ (with $\phi=0$ corresponding to the Be star in front of its companion). The interstellar absorption is fixed to $N_{\rm H}^{ISM}=2.7\times10^{20}$\,cm$^{-2}$ \citep{naz22}. The errors correspond to 1$\sigma$ errors, with the largest values provided if asymmetric. In the models of the top part of the Table, $kT_1$=0.085\,keV, $kT_2$=0.55\,keV, and $kT_4$=9.1\,keV while for the bottom part $kT_1$=0.085\,keV, $kT_2$=0.55\,keV, and $kT_3=kT_4$=9.05\,keV. 
 }
\end{table*}

\begin{table*}
  \centering
  \caption{Similar to Table \ref{fit} but for \ch\ and eROSITA spectra.  }
      \label{fit2}
 \begin{tabular}{lcccccccc}
 \hline\hline
 ID & $HJD$ & $\phi$ & $N_{\rm H}$ & $norm$ & $\chi^2$/dof & $F_{\rm X}^{obs}$& \multicolumn{2}{c}{Ct Rate (cts\,s$^{-1}$)}\\
 & $-2.45e6$ & & ($10^{22}$\,cm$^{-2}$) & ($10^{-2}$\,cm$^{-5}$) & & (2--10\,keV) & 2.--5.\,keV & 5.--8.\,keV\\
 \hline
26239      &9573.664 &0.63 & 15.7$\pm$2.0 & 2.38$\pm$0.24 & 30.80/36 & 1.63$\pm$0.10  \\
eROSITA1--4& -       & -   & 13.5$\pm$2.1 & 1.64$\pm$0.31 & 22.55/27 & 1.19$\pm$0.16  \\
eROSITA1   &8935.809 &0.84 & 13.3 (10.2--18.7) &2.77 (1.79--4.45) &21.69/22 & 2.03$\pm$0.70 & 0.42$\pm$0.08 & 0.16$\pm$0.07\\
eROSITA2   &9120.731 &0.23 & 22.2 (14.6--35.7) &1.78 (1.00--3.04) &         & 1.04$\pm$0.38 & 0.22$\pm$0.05 & 0.11$\pm$0.05\\
eROSITA3   &9297.519 &0.56 & 13.0 (6.7--17.9)  &2.23 (0.99--3.28) &         & 1.65$\pm$0.49 & 0.46$\pm$0.08 & 0.18$\pm$0.07\\
eROSITA4   &9480.901 &0.93 & 46.0 (30.8--66.6) &4.64 (2.49--7.88) &         & 1.71$\pm$0.69 & 0.09$\pm$0.03 & 0.20$\pm$0.06\\
 \hline 
 \end{tabular}        
\tablefoot{The absorbed thermal model has the form $phabs(ISM)\times phabs\times apec$ with $N_{\rm H}^{ISM}=2.7\times10^{20}$\,cm$^{-2}$ and $kT=9.1$\,keV. Fluxes are provided in units $10^{-11}$\,erg\,cm$^{-2}$\,s$^{-1}$. For individual eROSITA spectra, the errors on absorbing column and normalization factors were very asymmetric, hence the 1$\sigma$ confidence interval is provided instead. }
\end{table*}

\subsection{Spectra}

As can be seen in Fig. \ref{xmms}, the spectrum of \zt\ is dominated by hard X-rays. This emission comes from a very hot plasma, as the intense iron lines at 6.7--7.0\,keV testify. However, this emission suffers from a strong absorption that removes most flux below 4.\,keV. At the lowest energies, the flux decrease stops and a faint soft component starts to appear. It could not be seen in the previous \ch\ and eROSITA data due to their low signal-to-noise at these energies.

Comparing the two \xmm\ exposures reveals that the spectrum remains the same at the lowest and highest energies, but not in between (Fig. \ref{xmms}). It is only between 1.5 and 4.\,keV that changes are seen. At first sight, this seems to suggest a change in the absorption toward the hottest plasma. To ascertain what exactly happens, the spectra were fitted for $E>0.3$\,keV, within Xspec v.12.11.1, using solar abundances of \citet{asp09}. 

The first fitting trial considered a single thermal component ($apec$) absorbed by the interstellar medium ($N_{\rm H}^{ISM}=2.7\times10^{20}$\,cm$^{-2}$, see \citealt{naz22}) and some local material. While this fits well the high-energy part, the spectrum at low energies is not at all reproduced. We therefore introduced a second, cooler, component. This was still insufficient, so a third thermal component was added. This improved the situation at low energies although some small features near 1\,keV were still not perfectly reproduced, suggesting the need of a fourth thermal component. In addition, a Gaussian was clearly needed to reproduce the iron fluorescence line at 6.4\,keV (the line energy was fixed to that value and the line width was set to zero as the line is not resolved by the EPIC instruments). Adding a separate absorption for the warm thermal components resulted in an absorbing column compatible with zero, so it was discarded. Our final model thus consisted in four thermal components, plus one Gaussian and some absorption for the hottest component: $phabs(ISM)\times[apec_1+apec_2+apec_3+phabs\times apec_4+Gaussian]$. 

While the fitting used all EPIC spectra simultaneously, the two observations were fitted independently. As the first, second, and last best-fit temperatures agreed well, within errors, they were fixed and the fit was re-done with $kT_1$=0.085\,keV, $kT_2$=0.55\,keV, and $kT_4$=9.10\,keV. The resulting parameters are provided in Table \ref{fit}. A last trial was inspired by the analysis of the \xmm\ spectra of \gc\ \citep{smi04,rau22}: it considers $kT_3=kT_4$; in other words, only part of the hot component suffers from a (large) local absorption. The temperatures converged to similar values in both observations and thus were fixed ($kT_1$=0.085\,keV, $kT_2$=0.55\,keV, and $kT_3=kT_4$=9.05\,keV). The results of this trial are provided at the bottom of Table \ref{fit}. We note that most of the hot plasma is strongly absorbed. The normalization factor of $apec_4$ is at least 99.7\% of the full normalization of the hot component ($apec_3+apec_4$). Using a formal model of partial covering absorption (i.e. $phabs(ISM)\times[apec_1+apec_2+pcfabs\times apec_3+Gaussian]$) with the same temperatures confirms these results: the best-fit covering fraction is 1, with an error of 0.001.

Overall, both model choices provide fits of similar quality, with the parameters agreeing well. Comparing the two observations, \zt\ appeared only slightly (4\%) brighter in the first \xmm\ observation. One major change appears obvious, however: the near halving of the local absorbing column toward the hot component in the second observation (see also Fig. \ref{xmms}), which results in a large (+150\%) variation in the flux in the soft band.

\begin{figure}
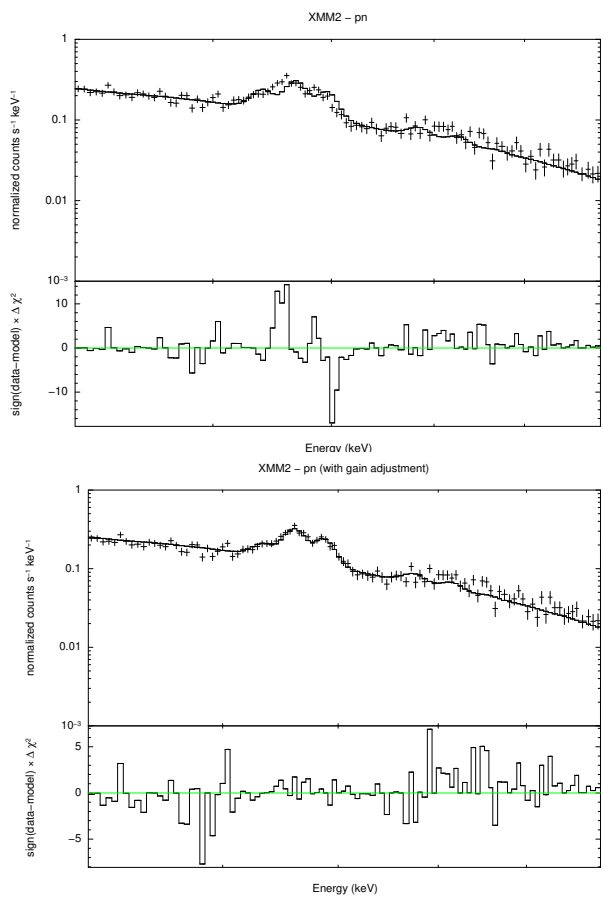

  \centering
  \includegraphics[height=8cm,angle=270,bb=40 0 560 700, clip]{iron_bpn.ps}
  \includegraphics[height=8cm,angle=270]{iron_bpn_withgainflimited.ps}
  \caption{Zoom on the 5.--10.\,keV interval, showing the iron complex recorded by EPIC-pn in the second \xmm\ observation, with the best-fit model superimposed. The top panel corresponds to the normal fit, while the bottom one makes uses of the response adjustement.}
  \label{iron}
\end{figure}

A specific feature in these X-ray spectra is the presence of the iron fluorescence line at 6.4\,keV, a typical characteristic of the \gc\ phenomenon. Examining the fitting, we find that there appears to be a shift of the feature toward lower energies compared to the models. It is particularly detectable for the pn data of the second \xmm\ observation (Fig. \ref{iron}), but a slight shift may also exist for MOS2 and the other pn spectra. This probably comes from imperfections in the charge transfer efficiency calibration. To solve this issue, the \xmm\ EPIC-pn team recommended that we allow for slight adjustments in the response matrices using the ``gain fit'' command under xspec. However, this solution remains approximate and therefore should be used with caution. In particular, the fitting should be restricted to a narrower energy band than the full energy range (we use 5.0--10.0\,keV) and with an offset of the gain kept within $\pm$10\,eV. Three models were used. The first one was a simplified version of the above model; that is, $apec+gaussian$ with the temperature fixed to 9.1\,keV and the line center and width fixed to 6.4 and 0.\,keV, respectively. The second and third models considered a continuum emission combined with three Gaussians of null intrinsic width centered on 6.4, 6.7, and 6.97\,keV, as was done for \gc\ \citep{rau22}. The continuum was either a bremsstrahlung with temperature allowed to vary (second model) or a power law (third model). The EPIC spectra were fitted all simultaneously or individually, to assess the errors. All three modeling methods yielded similar results. The gain adjustment, used for MOS2 and pn spectra, always resulted in a lower strength for the fluorescence line: the average equivalent width ($EW$) of the fluorescence line decreased from 72$\pm$18 to 59$\pm$11\,eV for the first \xmm\ observation after response adjustment and from 89$\pm$14 to 65$\pm$11\,eV for the second \xmm\ observation. In any case, the $EW$s appear similar in both observations as well as similar to or slightly higher than the fluorescence values recorded for \gc\ itself \citep{rau22}.

Finally, spectra were extracted before and during the shallow ``softness dip'' detected at the end of the first \xmm\ observation (i.e. $HJD<$ or $>2\,460\,024.1$, see Section 3.1). We performed a spectral fitting similar as for the full observation. We allowed however the absorbing column toward the warm plasma components to vary ($phabs(ISM)\times[phabs\times (apec_1+apec_2+apec_3)+phabs\times apec_4+Gaussian]$, with $kT_1$=0.085\,keV, $kT_2$=0.55\,keV, and $kT_4$=9.10\,keV). This yielded inconclusive results, since parameters agreed within errors. A trial was then made to fit both subexposure spectra together, letting absorptions vary but allowing only for a global scaling of the normalization factors between the two observations. The resulting 1-$\sigma$ confidence intervals for the absorbing columns toward the warm plasma components are $0.-0.10$ and $0.09-0.24\times 10^{22}$\,cm$^{-2}$ before and during the dip, respectively. This shallow softness dip thus appears consistent with a marginal increase in absorption, but it is certainly not a drastic change. 

To complement the \xmm\ results, the \ch\ \citep{naz22} and eROSITA \citep{naz23} spectra were fitted again. This time, we fixed the hot temperature to the same value as for \xmm\ and did not consider warm plasma components since they cannot be detected in these lower quality spectra (Table \ref{fit2}). The fitting of the \ch\ observation indicated a large absorbing column, which agreed with that of the first \xmm\ exposure, within errors. The absorption found for the average eROSITA spectrum appears intermediate, at 2$\sigma$ of each of the \xmm\ values. For eROSITA observations, we further investigated the spectral shape in each of the four survey semesters. These individual exposures are short (200\,s) but each dataset consists of several scans taken over $\sim$1\,d (i.e. at the same orbital phase). They thus enabled us to derive further constraints on orbital variations (if any). As in \citet{naz23}, the fitting to eROSITA data was performed with a power law to constrain the optical loading contribution (dominating below 1\,keV) in addition to the absorbed thermal component. We considered the four individual spectra together, allowing only for local absorptions and normalization factors to vary freely while forcing other parameters to keep the same values in the four individual fits. This allowed us to better constrain the optical loading, which should remain constant (within errors) between exposures. Results are provided in Table \ref{fit2}. It may be noted that similar results are achieved (1) if the optical loading power law has parameters fixed to those found for the best-fit to the average eROSITA spectrum and (2) if the spectral bins below 1\,keV are ignored and the remaining spectral bins are fitted with a simple thermal component. In the \xmm\ datasets, the largest variations appear in the 2.--4.\,keV energy band. This can be confirmed by eROSITA observations, as the survey data analysis provides equivalent on-axis, full-PSF count rates in the 2.--5.\,keV and 5.--8.\,keV energy bands (Table \ref{fit2}). These count rates confirm the behavior observed for \xmm, with similar values, within errors, at the highest energies and larger variations in the medium energy range. This is directly reflected in the spectral fitting by larger absorbing columns when the 2.--5.\,keV count rates are lower.

\zt\ (type B1IVe) is an early B-type star. The presence of the warm plasma components in its X-ray spectrum could simply reflect the typical X-ray emission of massive stars. To test this idea, we evaluated the unabsorbed (i.e. corrected for interstellar absorption) flux of these components in the 0.5--10.\,keV band. In our first model, this intrinsic flux would correspond to the sum of the fluxes of the first three $apec$ components, even if the last one has a rather high temperature for such intrinsic X-rays. This flux amounts to 7.3 and $8\times10^{-14}$\,erg\,cm$^{-2}$\,s$^{-1}$ for the first and second \xmm\ observations, respectively. For the second model, the intrinsic flux would correspond to the sum of the fluxes of the first two $apec$ components; that is, 2.8 and $2.5\times10^{-14}$\,erg\,cm$^{-2}$\,s$^{-1}$ for the first and second \xmm\ observations, respectively. Given the distance and the bolometric luminosity of \zt\ (136\,pc and $\log(L_{\rm BOL}/L_{\odot})=3.75$, see \citealt{naz22} and references therein\footnote{This bolometric luminosity is different from that used in \citet{naz23}. This luminosity difference mostly comes from a different choice of bolometric correction. It was based on the effective temperature of \zt\ in \citet{naz22} and on the spectral type of \zt\ in \citet{naz23}. However, the effective temperature of \zt\ appears lower than usual for its spectral type.}), these fluxes convert into X-ray luminosities of 0.6--1.8$\times10^{29}$\,erg\,s$^{-1}$; hence, a luminosity ratio $\log(L_{\rm X}/L_{\rm BOL})$ between --8.6 and --8.1. Such values seem reasonable for this type of star. They appear at the faint end of what was found for Be stars in X-rays \citep{naz18,naz23}. The faint and soft X-ray emission is therefore most probably intrinsic to the Be star itself. In comparison, we note that the full X-ray luminosity of \zt, after correcting for interstellar absorption, is much larger, around $4\times10^{31}$\,erg\,s$^{-1}$.

\section{Discussion}

\subsection{The variations in \zt}

The X-ray observations indicate the presence of both warm and hot plasma, with the latter component suffering from a very strong local absorption. This strong absorption appears unrelated to the ``softness absorption dips'' apparently occurring randomly in time, detected for \gc\ \citep{ham16,rau22} and possibly toward the end of the first \xmm\ observation of \zt. Such dips affect the very soft energy range (below 1.5\,keV) and correspond to increased absorption (up to $10^{22}$\,cm$^{-2}$) toward the warm+hot plasma components. In \zt, the main absorption is large (at least $10^{23}$\,cm$^{-2}$), impacts only the hottest plasma component, and affects the flux in the medium energy range (1.5--4.\,keV, see Fig. \ref{xmms}). The strong absorption therefore represents a different, additional, feature from the softness dips. In this context, it is important to note that this strong absorption is detected in all X-ray observations, taken by independent facilities (\ch, eROSITA, \xmm) at different epochs; hence, it is not a transient feature observed only once.

\begin{figure}
  \centering
  \includegraphics[width=8.5cm,bb=19 145 590 420, clip]{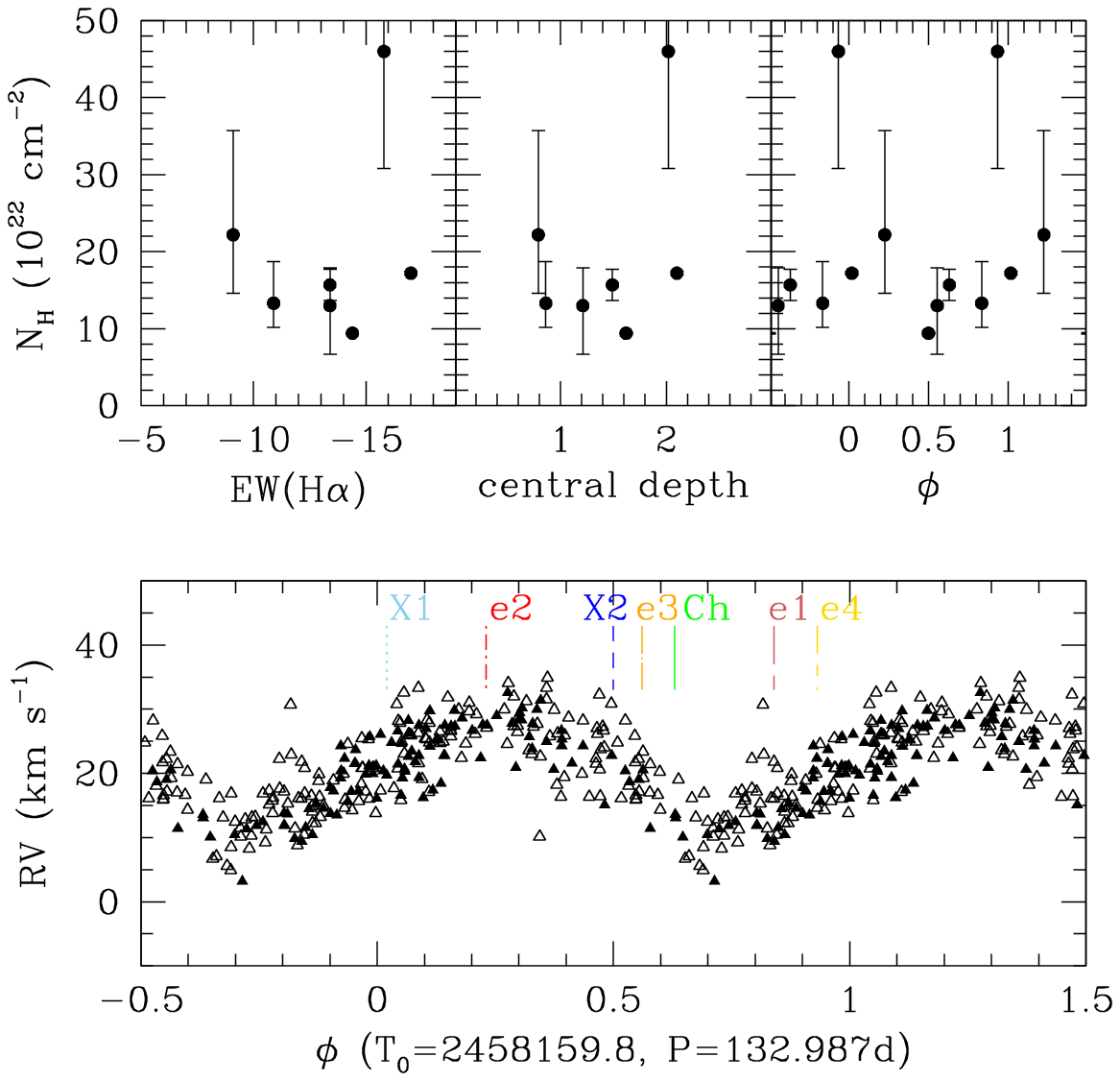}
  \includegraphics[width=8.5cm]{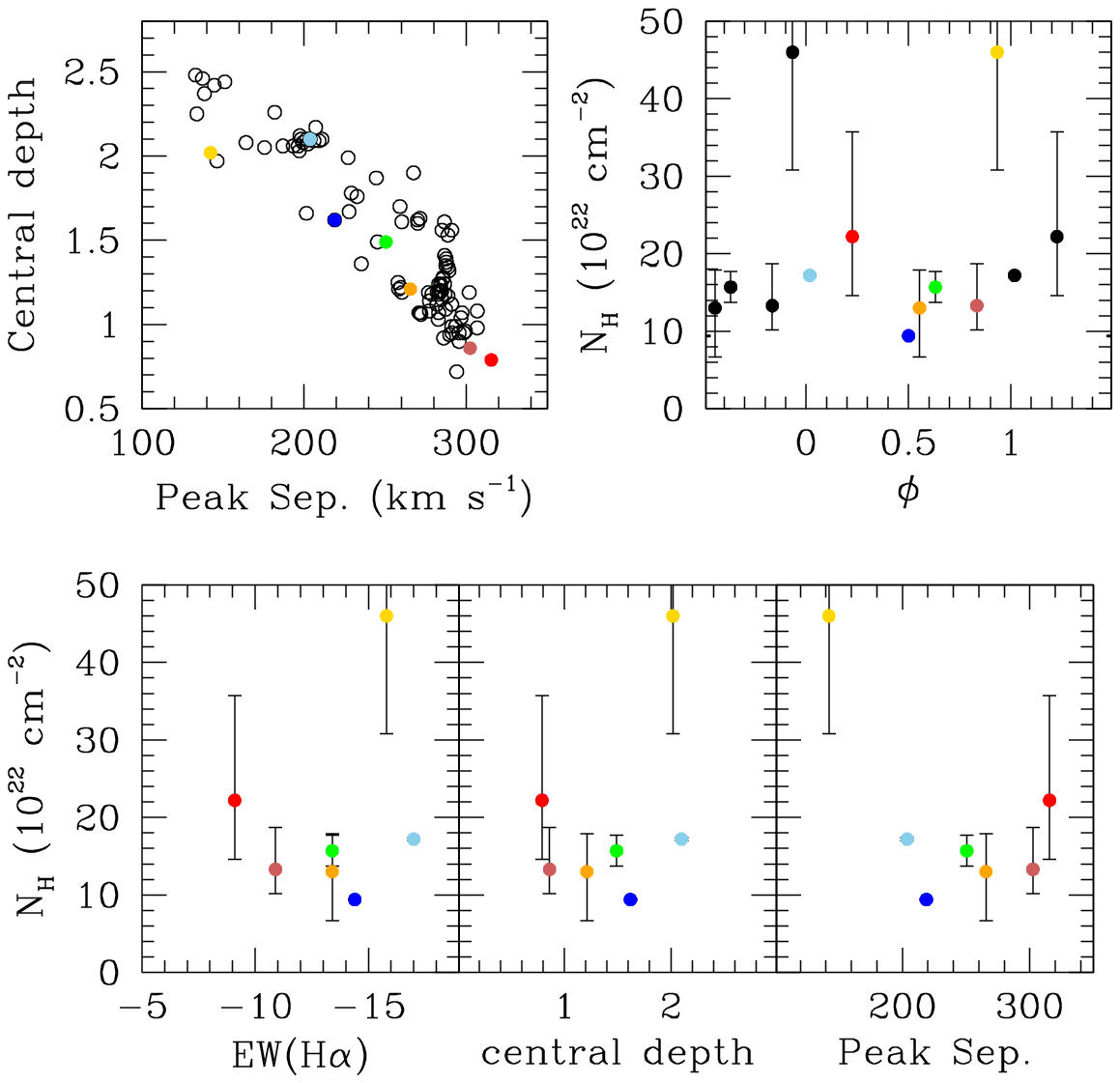}
  \caption{Comparison of optical and X-ray characteristics. {\it Top panel:} Radial velocities of the Be star in \zt, measured from the H$\alpha$ line using the double Gaussian method (filled triangles from \citealt{naz24} and empty triangles from the appendix of \citealt{naz22}) and phased with the ephemeris of \citet{naz22}. Vertical lines show when X-ray observations were taken. Orbital phases 0 and 0.5 correspond to conjunctions with the Be star and its companion in front, respectively, while phases 0.25 and 0.75 correspond to quadratures. {\it Bottom panels:} Comparison between the X-ray absorbing column and various diagnostics: orbital phase as well as $EW$, central absorption depth, and peak separation of the H$\alpha$ line. The colors are the same as in the top panel. }
  \label{opt2a}
\end{figure}

An additional characteristic of the strong absorption is its variability, with a doubling of the absorbing column between the two \xmm\ exposures. This is found whatever the spectral model choice, and hence is a very robust result. The origin of this variability can be discussed. The first thing that comes to mind is of course an orbital effect. The first \xmm\ observation was taken close to conjunction, when the Be star was in front of its companion ($\phi=0.02$), while the second one, with a lower absorbing column, was taken at the opposite phase, when the companion was in front of the Be star ($\phi=0.50$, Fig. \ref{opt2a}). One would off-hand think that the X-ray absorption increases as the Be star and its disk enter the line of sight toward an X-ray emitting region attached to the companion. We note in this context that the fast rotation of Be stars is thought to originate from a mass-transfer event, suggesting that the orbital and rotational axes should be nearly coincident. It is worth considering also that a Be disk inclined with respect to the orbit would lead to regular outbursts at companion's crossing in the case of accretion-powered X-rays (as seen in high-mass X-ray binaries), but none are seen for \zt\ or other \gc\ analogs. The orbital plane of the \zt\ binary can thus be considered to be similar to that of the Be disk.

\begin{figure}
  \centering
  \includegraphics[width=8.5cm]{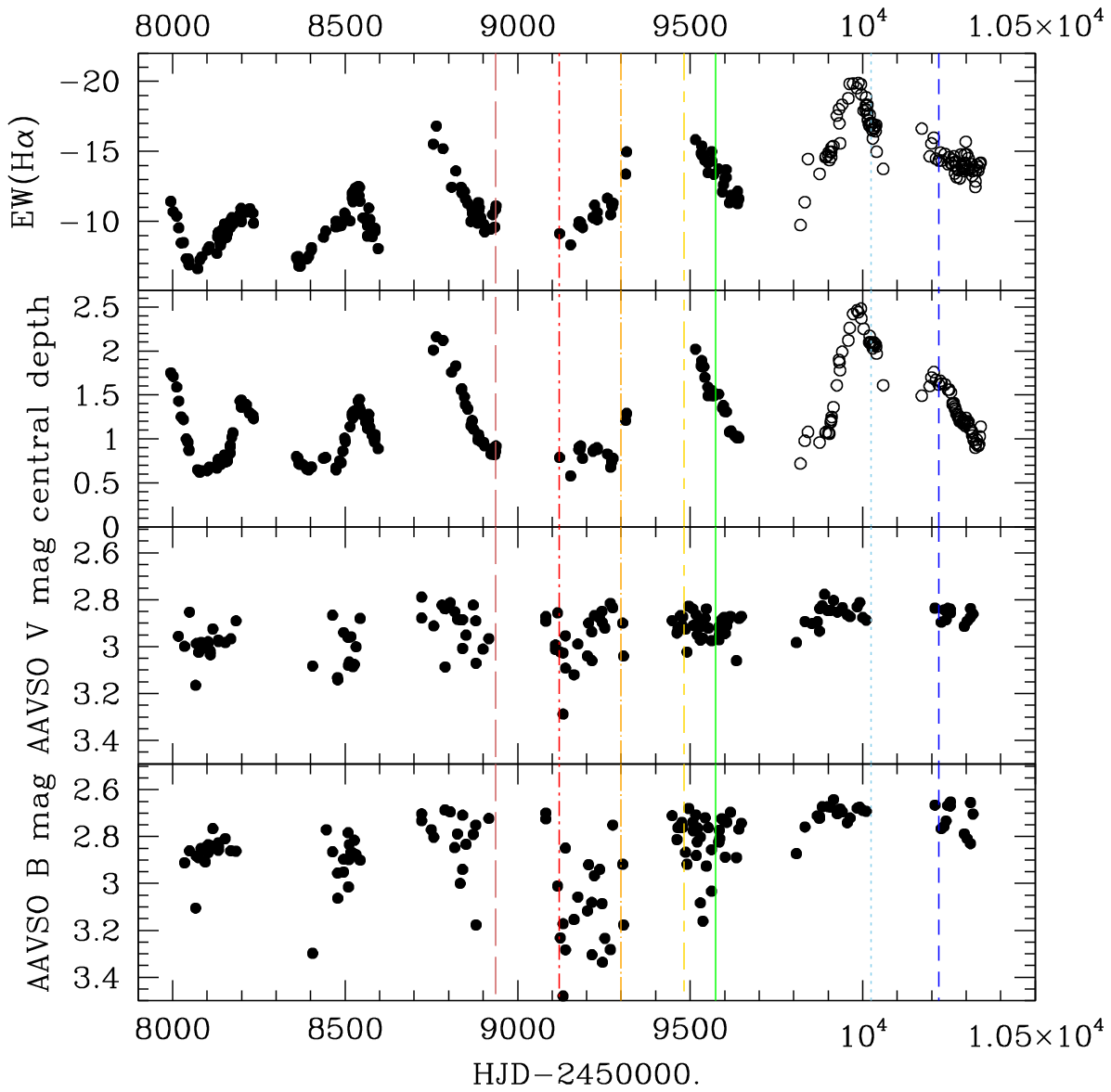}
  \includegraphics[width=8.5cm]{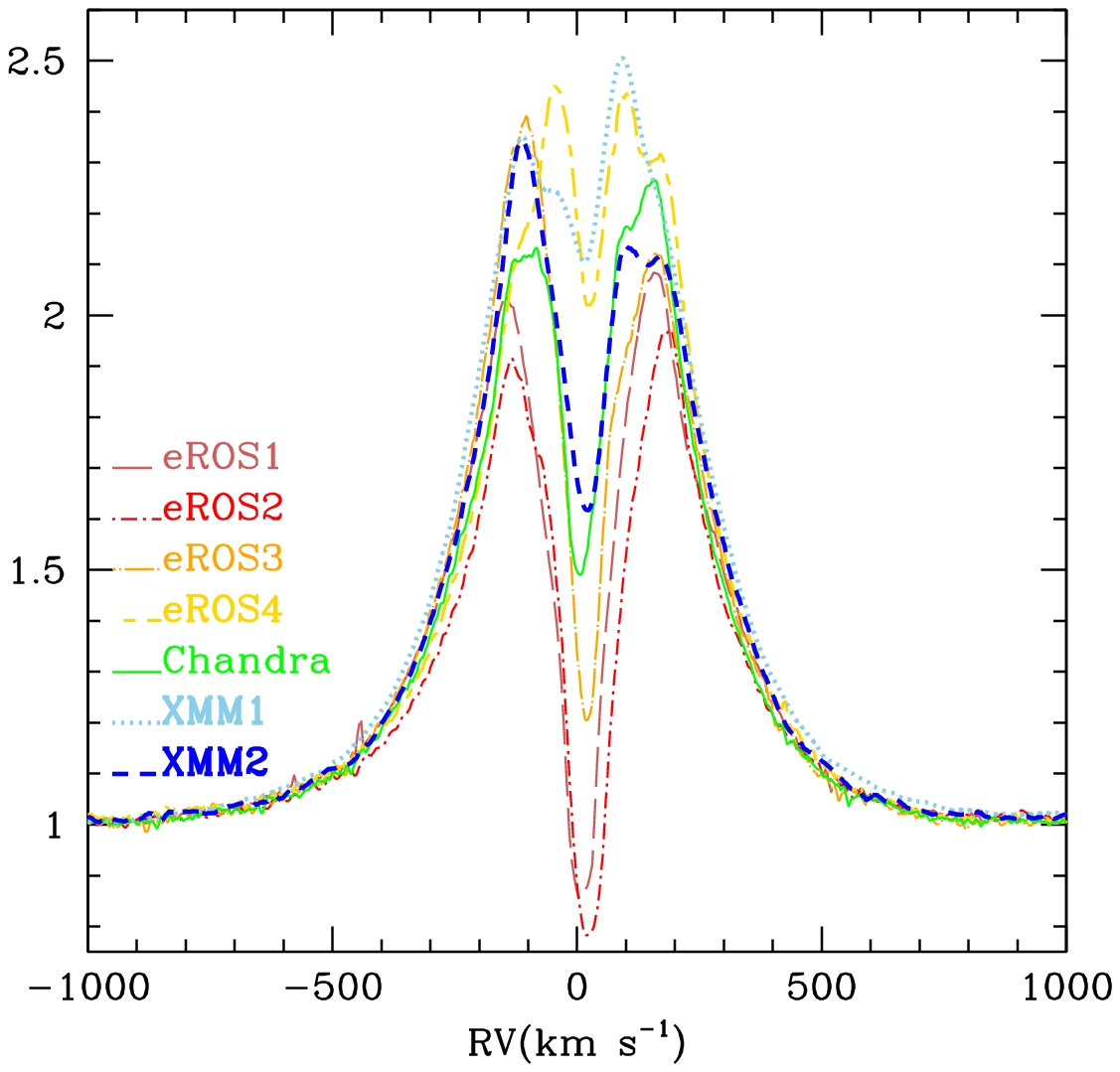}
  \caption{Optical behavior of \zt. {\it Top panels:} Evolution with time of $EW$ and central depth of the H$\alpha$ line from amateur spectra (BeSS database - black dots from \citealt{naz22}, circles from \citealt{naz24}), as well as $B,V$ magnitudes of \zt\ as reported in the AAVSO database. {\it Bottom panel:} H$\alpha$ line profiles at times closest to those of the X-ray observations.  }
  \label{opt2}
\end{figure}
 
Although the \ch\ data were gathered at a phase ($\phi=0.63$) just after that of the second \xmm\ observation, with the companion in front, they present an absorbing column close to that of the first \xmm\ observation, when the Be star was in front. In addition, the four eROSITA datasets were taken at various phases: before the first \xmm\ exposure for the first and fourth eROSITA observations ($\phi=0.84,0.93$), in between the second \xmm\ and single \ch\ exposures for the third eROSITA one ($\phi=0.56$), and at quadrature ($\phi=0.23$, between the two \xmm\ exposures) for the second eROSITA one (see top panel of Fig. \ref{opt2a}). However, their absorptions can hardly be reconciled with an orbital effect. Indeed, the first and third eROSITA spectra ($\phi=0.84,0.56$) share similar absorptions to those of the \ch\ and first \xmm\ spectra ($\phi=0.63,0.02$) despite all four observations having very different phases. Much larger absorptions are found in the second and fourth eROSITA datasets ($\phi=0.23,0.93$) although the latter one has a phase close to that of the first eROSITA scan ($\phi=0.84$). The middle right panel of Fig. \ref{opt2a} summarizes the variations in the absorbing column with respect to phase, graphically demonstrating the absence of the link that has just been described. At similar phases, very different absorptions can be observed. Therefore, a coherent, phase-locked absorption effect seems unlikely.

Another possible explanation for the absorption variability is a change in the disk. With this aim we examine in the top panel of Fig. \ref{opt2} the evolution of some common disk diagnostics. On the one hand, there does not seem to be drastic photometric changes between the times of the X-ray exposures, although the AAVSO\footnote{https://www.aavso.org/} $B,V$ magnitudes (reported by a single contributor, the University of Illinois) are quite noisy, and hence could hide small-to-moderate ($<0.2$\,mag) variations. On the other hand, the H$\alpha$ line is continuously varying. \citet{naz22} showed that the recent H$\alpha$ changes mostly consists of a varying central absorption, without much V/R asymmetry and without a specific recurrence timescale. In the last two optical observing seasons (2022--23 and 2023--24), the triple- or multiple-peaked profile returned when the central absorption was minimum, as did some slight V/R asymmetry at some epochs \citep{naz24}. However, the situation still appears far from the cyclic changes with strong asymmetries observed in 1997--2008 \citep{ste09}. The $EWs$ of the H$\alpha$ line, measured from --600\,\kms\ to +600\,\kms\ on the BeSS amateur spectra closest to the X-ray observations\footnote{See http://basebe.obspm.fr/basebe/ and the tables in \citet{naz22} and \citet{naz24}. The BeSS spectra closest to \xmm\ observations were taken on 2023 March 20 and October 3, respectively; in other words, the optical and X-ray data were taken less than a day apart. A snapshot covering only the H$\alpha$ line was taken five days before the \ch\ exposure (on 2021 December 20) but the closest echelle spectrum is dated from 11 days before (2021 December 13). For eROSITA data, optical spectra covering H$\alpha$ were taken within a day of the first two X-ray observations (2020 March 27 and September 29). The optical spectra closest to the third and fourth eROSITA observations were however taken 16 and 25 days later (2021 April 8 and October 28), respectively. Echelle spectra were obtained at even more distant dates.}, yield a range of values: --10.9, --9.1, --13.4, --15.8, --13.4, --17.0, and --14.4\,\AA\ for the four eROSITA observations, the single \ch\ exposure, and the two \xmm\ observations, respectively.

More precisely, when examining the line profiles (see bottom panel of Fig. \ref{opt2}), the wings of the H$\alpha$ line always remain similar, but both the asymetry and central absorption depth differ. At the epoch of the second \xmm\ and third eROSITA observations, the violet peak appears slightly dominant, while the red one is dominant for the first \xmm\ exposure and \ch\ dataset. The profile remained nearly symmetric at the time of the other eROSITA observations. Again, there is no obvious link with the X-ray absorption changes. In parallel, the first two eROSITA observations were taken when the central absorption in H$\alpha$ was very strong. The third eROSITA, single \ch, and second \xmm\ observations were acquired when that central absorption presented a medium depth, while the central absorption displayed small depth values during the last eROSITA and first \xmm\ observations. Other lines (such as H$\beta$, He\,{\sc i}\,$\lambda$6678\,\AA, Si\,{\sc ii}\,$\lambda$5321,5363\,\AA, N\,{\sc ii}\,$\lambda$6242--8\,\AA, and O\,{\sc i}\,$\lambda$7771--5\,\AA, using amateur echelle spectra) confirm these trends although their variations are much smaller (due to their different formation regions).

In contrast, the X-ray absorbing column is largest for the second and fourth eROSITA datasets, which lie at opposite extremes of the H$\alpha$ central absorption strength (minimum below the continuum level or at twice the continuum value, respectively, see Fig. \ref{opt2}). Also, the X-ray absorbing column appears the lowest in the second \xmm\ observation while a medium depth is then recorded for H$\alpha$. In fact, for the more precise \xmm\ exposures, the smallest central absorption depth in H$\alpha$ is detected in the first \xmm\ exposure, when the X-ray spectra presented the largest(!) absorbing column. Again, no obvious link between the two absorptions can be seen (see bottom panels of Fig. \ref{opt2a}). In recent years, the peak separations in the H$\alpha$ line profiles appear anticorrelated with the strengths of the central absorption (i.e. deeper central absorptions are generally associated with more separated peaks - see \citealt{naz24}). The optical spectra taken at the phases closest to the X-ray observations follow this trend (Fig. \ref{opt2a}). As for the central absorption depth, there is no clear relationship between peak separation and X-ray absorption (Fig. \ref{opt2a}). Therefore, (1) there is no drastic global change in the Be disk (i.e. build-up or disappearance), as traced by the $B,V$ magnitudes and the optical spectra, between the different epochs of X-ray observations and (2) the detailed H$\alpha$ profile properties (depth of the central absorption, symmetry, peak separation, and $EW$) appear uncorrelated with the X-ray absorbing column values. A direct link between the Be disk properties and the X-ray absorption thus appears unlikely. 

In \zt, the X-ray absorption changes therefore seem unrelated to the orbital phase or to the disk variations detected in the H$\alpha$ line. This is not unexpected for \gc\ objects. Indeed, no evidence of a phase-locked behavior was found in the (only) two previously thoroughly monitored cases, \gc\ \citep{smi12b,rau22} and $\pi$\,Aqr \citep{naz19}. In addition, the link between the X-ray emission and the H$\alpha$ properties appears far from obvious in previous studies, with some objects keeping their \gc\ peculiarities while displaying very little H$\alpha$ emission \citep{naz22mon}. This implies that the line of sight toward the hot plasma is different from the line of sight toward the companion (and its local environment) or from that in the disk (as traced by H$\alpha$). In this context, it is also useful to note that the $EW$ of the fluorescence Fe\,K$\alpha$ X-ray line remains rather stable in the datasets of both \gc\ \citep{rau22} and \zt\ (this work).

\subsection{\zt\ in context: The star-disk interaction scenario}
With its high temperature, high luminosity, strong iron fluorescence line, and short-term variations, \zt\ clearly displays all characteristics of the \gc\ analogs. Its particularity is the presence of a strong and varying absorbing column toward the emission by the hottest plasma. To underline this uniqueness of \zt, one may compare with the detailed spectral fitting of \gc. \citet{rau22} used a combination of absorbed warm and hot plasma components, as above. They only found large absorptions toward the hottest component, up to $2\times 10^{23}$\,cm$^{-2}$, if such absorptions applied to a small fraction ($<$20\%) of the hot component emission. When that fraction was larger, notably when it reaches over 90\% as found in \zt\ (see Sect. 3.2), the column appeared much more modest (up to $1.4\times 10^{22}$\,cm$^{-2}$). The X-ray absorption of \zt\ therefore truly appears exceptionally high.

\begin{figure}
  \centering
  \includegraphics[width=9cm]{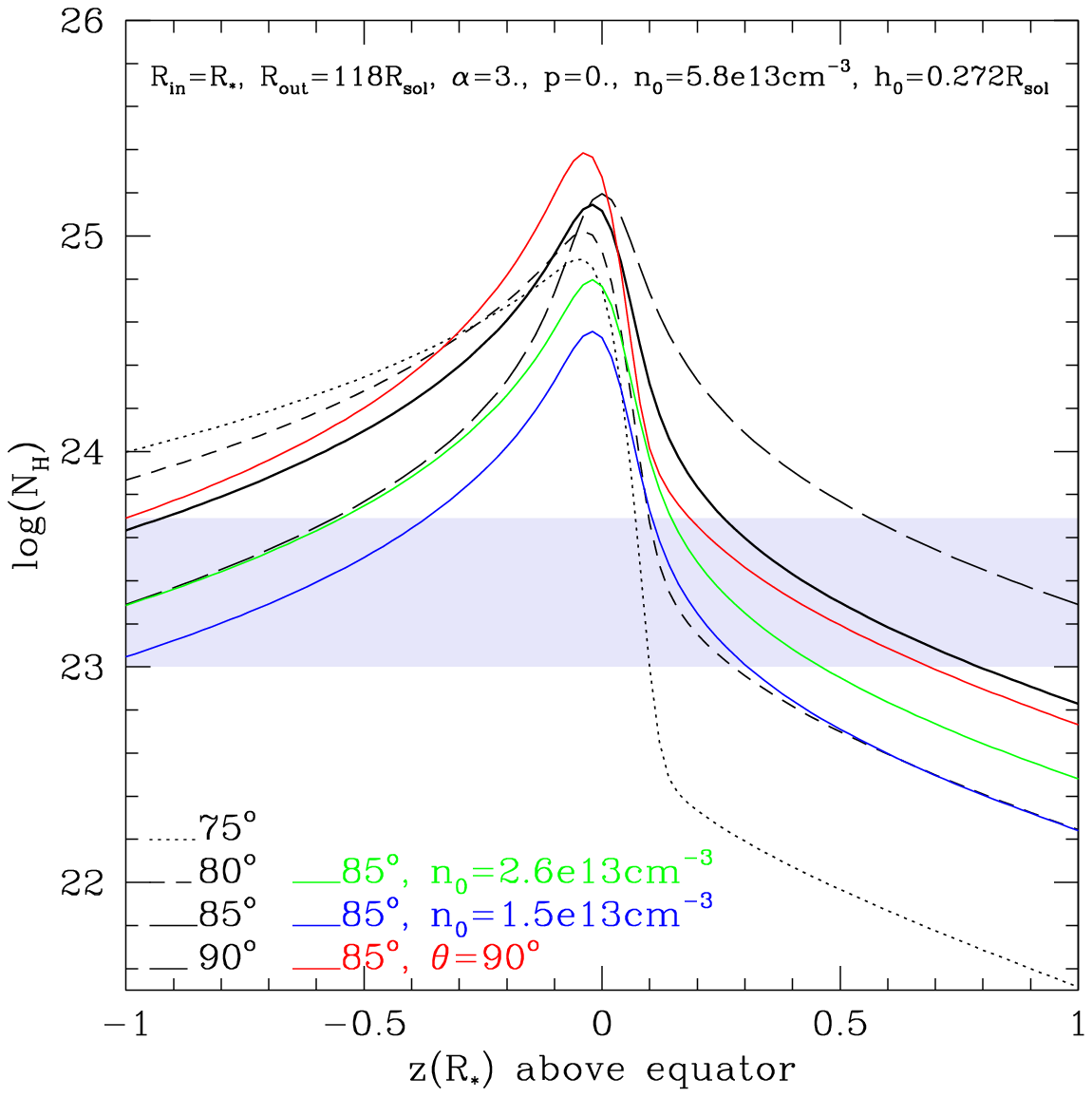}
  \includegraphics[width=9cm]{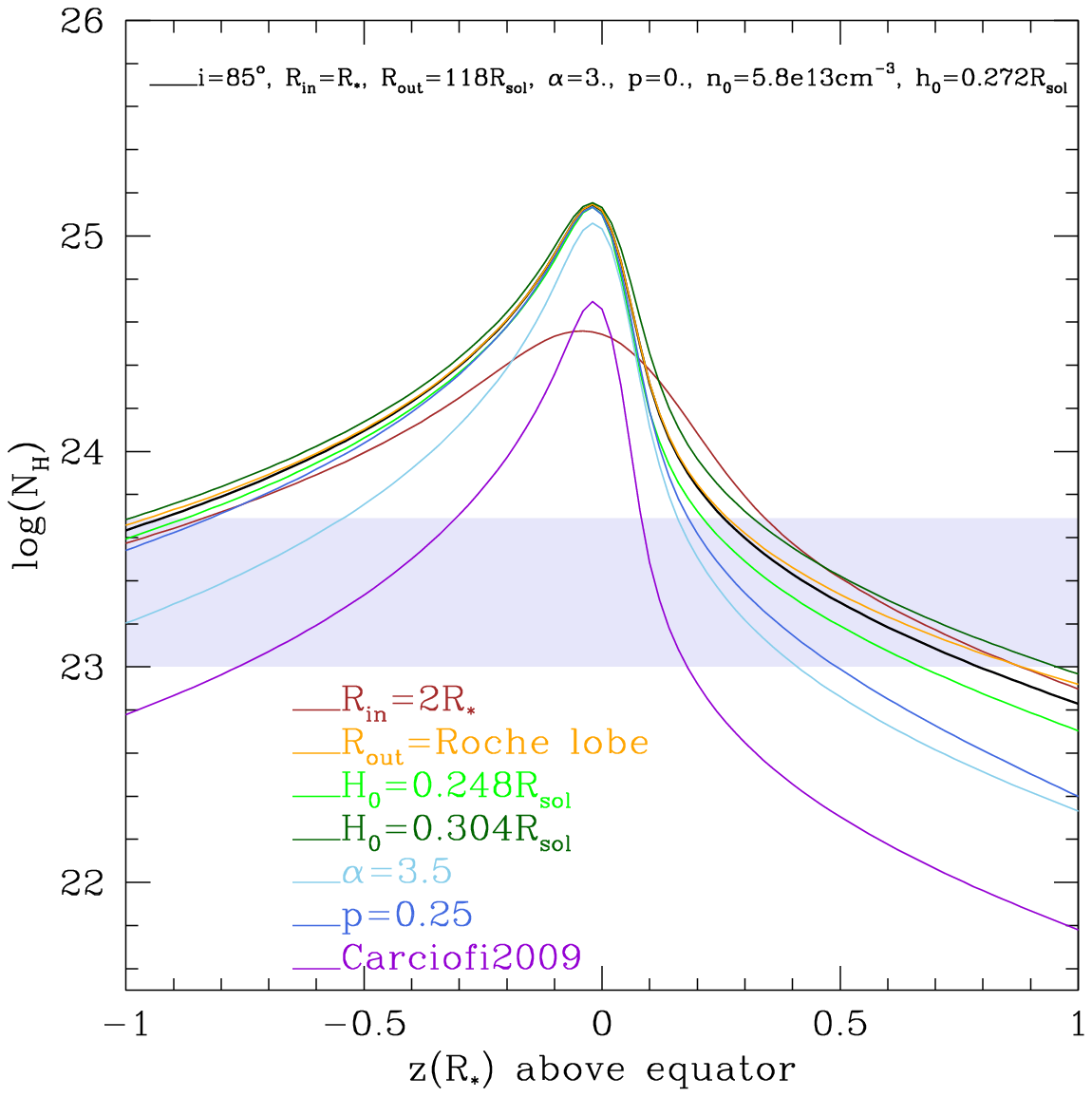}
  \caption{Evolution of the predicted absorbing column from disk material as a function of position above the equatorial plane $z$ and various parameters (the baseline parameters are noted in black and the thick black line provides the associated column; $\theta=0^{\circ}$ everywhere unless otherwise stated). The observed range of columns is shown by the shaded light lavender area. }
  \label{nh}
\end{figure}
 
However, it may be noted that, among the \gc\ analogs with known inclinations, \zt\ is also the only one with the Be disk seen edge-on. Most probably, high absorption and high inclination are related. In the context of the magnetic star-disk interaction scenario, hard X-rays arise from regions close to the star and the disk. This implies that they may suffer from absorption due to the Be disk. To assess whether this is in line with observations, we calculated the absorbing column due to the Be disk material along the line of sight. To this aim we considered the usual density dependencies for Be disks:\\ $n(r,Z)=n_0(R\*/r)^{\alpha}e^{-Z^2/(2H^2)}$ with $r$ the distance to the star's center in the equatorial plane, $Z$ the distance above the equatorial plane, and the scale height $H=H_0(r/R_*)^{(3-p)/2}$.

Several sets of parameters were considered for these calculations, inspired by \citet{rau24}. The equatorial radius of the star was set to 7.7\,R$_{\odot}$ and the density at the inner edge of disk $1.3\times10^{-10}$\,g\,cm$^{-3}$ (corresponding to $n_0=5.8\times10^{13}$\,cm$^{-3}$ since the absorbing column $N_{\rm H}$ concerns only hydrogen). The latter value is an average based on published densities \citep{gie07,car09,tou11}, but values of $n_0=1.5$ and $2.6\times10^{13}$\,cm$^{-3}$ were also tried. Indeed, it may be noted that the H$\alpha$ line has recovered from its 2013 minimum. The maximum amplitude is about 2.5 times the normalized continuum and the $EW$ was around --15\,\AA\ at the time of the \ch\ and \xmm\ observations (Fig. \ref{opt2}). This can also be compared to values of about 3 times the continuum and around --20\,\AA\ during the well-studied $V/R$ cycles of 1997--2008 \citep{ste09,pol17}. Therefore, a somewhat lower density may be expected. The exponents $\alpha$ and $p$ were assumed to be 3 and 0 by default, respectively, but values of 3.5 and 0.25 were also considered. Inclinations of 75, 80, 85, and 90$^{\circ}$ were examined (with 85$^{\circ}$ the default value). The disk was assumed to extend either to the Roche lobe radius (149\,R$_{\odot}$) or to the 3:1 truncation radius (118\,R$_{\odot}$). The emitting point, at which the integration starts, was set at a distance $R_{in}$ from the star's center and at a distance $z$ above the equatorial plane. Values for $R_{in}$ were $R_*$ (at the photosphere, default case) or $2R_*$ and extreme $z$ values are $\pm R_*$. The longitude of the emitting point was allowed to vary between $\theta=0^{\circ}$ (i.e. emission arising on the meridian facing the observer) and 90$^{\circ}$ (i.e. perpendicular to the meridian). The initial scale height is given by $H_0=c_S(R_*)R_*/V_{Kep}$, with $V_{Kep}$ the Kepler rotational velocity at the photosphere ($V_{Kep}=\sqrt{GM_*/R_*}$) and $c_S$ the sound speed; that is, $c_S=\sqrt{\gamma kT/(\mu m_{\rm H})}$. Depending on the choice of $\gamma$ (1 if isothermal, 5/3 if adiabatic) and temperature (2/3, 0.8, or 1.0 times the effective temperature $T_{eff}$ of 19300\,K), various $H_0$ values are found; we used 0.272, 0.248, and 0.304\,R$_{\odot}$. Finally, the parameters of \citet{car09} were considered: $R_{out}$=130\,R$_{\odot}$, $R_{in}=R_*$, $n_0=2.6\times10^{13}$\,cm$^{-3}$, $H_0=0.230$\,R$_{\odot}$ (from $\gamma=1$ and $T=18000$\,K), $i=85^{\circ}$, $\alpha=3.5$ and $p=0$.

Figure \ref{nh} graphically shows the predicted columns. As was expected, the columns are larger if the emitting point is closer to the equatorial plane and ``below'' the disk (seen from the observer's point of view, corresponding to $z<0$). The ``above'' and ``below'' difference is of course larger for lower inclinations. Longitude has only a small impact on the predicted column, and changing the outer radius in a reasonable range has no significant consequence. Enlarging the initial radius $R_{in}$ decreases the maximum, but has less impact if emission occurs outside of the equatorial plane. Absorptions for such emissions outside the plane are much more sensitive to the density exponent $\alpha$. Finally, modifying the initial density $n_0$ produces a simple global scaling of the derived absorbing column. Overall, a deeply embedded emitting region - in other words, near the photosphere and on the equatorial plane - would suffer from very large (more than one dex!) absorptions compared to those derived from the X-ray fitting ($N_{\rm H}=1-5\times10^{23}$\,cm$^{-2}$, see Tables \ref{fit} and \ref{fit2}). However, the derived absorbing columns appear similar to the observed ones for moderate elevations above the equatorial plane. The observed columns are thus not incompatible with absorption by the Be disk.

Finally, it may be noted that the absorbing column toward equatorial emission remains large for putative \zt-like systems with low inclinations: $0.5-2\times10^{24}$\,cm$^{-2}$ for i=20$^{\circ}$ vs $0.4-1.6\times10^{25}$\,cm$^{-2}$ for i=90$^{\circ}$, depending on the choice of disk parameters (see above). As absorbing columns toward other known \gc\ analogs are lower, this suggests that either X-ray emissions always arise outside of the equatorial plane of the Be disk, or that the Be disk model is inadequate in regions very close to the star. Also, the lack of correlations with spectroscopic features linked to the Be disk remains puzzling. Unfortunately, no precise long-term X-ray and visible photometry is available to make a comparison as for \gc\ \citep{mot15,rau22}. Clearly, detailed MHD modeling of this scenario would be needed to clarify the emission processes.

\subsection{\zt\ in context: The accreting white dwarf scenario}

White dwarfs have been proposed as the sources of the X-ray emission of \gc\ analogs \citep{ham16,gie23}. In this subsection, we take a look at their properties and compare them to those recorded for \zt. Up to now, only a few cases of WDs paired with Be stars have been reported in literature and their X-ray emissions are very bright and supersoft \citep{kah06,ori10,li12,stu12,cra18,coe20,ken21}, in stark contrast with the high-energy properties of \zt\ and other \gc\ analogs.

Other WD systems, in particular the cataclysmic variables (CVs), may appear hard at X-ray wavelengths, though (for a thorough review, see \citealt{muk17}). In these short-period ($<2$\,d) binaries, mass is transferred from a late-type star to the WD through the L1 Lagrangian point. The material can then form an accretion disk (for non-magnetic cases, such as dwarf novae), fall directly on the WD surface (for the highly magnetic cases called polars), or fill a truncated disk and then fall into the WD following the magnetic field lines (for intermediate polars, IPs). Symbiotic systems may also emit hard X-rays but the configuration is different: the companion of the WD is a giant star, typical orbital periods are longer and the accretion is fueled by winds (although disks have been detected in some cases).

Unfortunately, no identified accreting WD system displays a geometry close to that expected for \gc\ analogs in general and \zt\ in particular. In addition to its long orbital period, the decretion disk, possibly truncated at the 3:1 resonance (118\,R$_{\odot}$ for \zt, see above), lies well inside the Roche lobe of the Be star. The flow of matter entering the WD Roche lobe would then be concentrated along the plane of the decretion disk with velocities dominated by the Keplerian tangential component. Such an accretion geometry and velocity pattern contrast with those in CVs and symbiotic stars. In CVs, the mass donor, companion to the WD, fills its Roche lobe and matter enters the WD Roche lobe through L1 while most symbiotic stars are fed by winds flowing from the giant companion star. A comparison nevertheless remains interesting, to assess the plausibility of the accreting WD scenario\footnote{Contrary to the case of neutron stars, the WD formation process is not expected to apply any signficant kick to the orbit. If tidal spin alignment occurred during the mass-transfer episode that spun up the former secondary star (and now Be star), the orbital plane will nearly match that of the decretion disk and that of any accretion disk created around the WD. We therefore will consider the system coplanar in all its components. }.

Regarding temperatures, there are usually two types of spectral components in X-rays from accreting WDs, a soft one (arising at or near the WD surface, at tens of eV) and a hot one arising in the accreting flow, with values well above 5\,keV and typically of several tens of keV. Because of the latter, an intense iron complex naturally arises in the X-ray spectrum. Since cooler material is present, such as the WD surface, a fluorescence line is also detected for these systems. Comparing the properties of the lines in the iron complex (fluorescent Fe\,K$\alpha$, Fe\,{\sc xxv}, Fe\,{\sc xxvi}), \zt\ appears to lie among accreting WDs, but between IPs and quiescent dwarf novae \citep{xu16}. 

Regarding variability, accreting WDs display several types, each with different characteristics. Over the long term, there are of course nova events, recurrent or not, resulting in X-ray outbursts. No outburst has been detected for \gc\ analogs up to now, however. Over the short term, there is also an aperiodic variability called ``flickering''. In both non-magnetic CVs (dwarf novae, \citealt{bal19}) and magnetic CVs (IPs, \citealt{rev11}), the power spectra of the X-ray light curves first follow $P\propto f^{-1}$ then $P\propto f^{-2}$ with a break near (a few) mHz for non-magnetic cases or (a few) 10mHz for magnetic cases. This dual slope is interpreted as reflecting density fluctuations in the outer or inner disk, respectively. These fluctuations are thought to be linked to variations in the accreted mass flow \citep{bal20}. When there is no disk, as in polars, the fluctuating accretion directly affects the X-ray emission. For example, in the polar AM\,Her, after taking out the orbital modulation, the X-ray light curve displays variations following $P\propto f^{-1.34}$ (\citealt{bea97}, see also Fig. 4 of \citealt{chr00}). In the extremely hard \zt, Figure \ref{four} above provides the periodogram amplitude - the squared root of power - so that $P=A^2\propto f^{-1.5}$ with no clear break between 1 and 10mHz. Due to noise (see Sect. 3.1 above), the exponent is not well defined, however, and $P=A^2\propto f^{-1}$ could also fit. A better characterization was done for the brighter \gc, revealing $P\propto f^{-1.04}$ at low frequencies and a break near 5\,mHz followed by a steeper trend $P\propto f^{-1.44}$ \citep{lop10}. Overall, the frequency behavior thus appears quite similar in all these sources. One must remain careful, however, as this does not necessarily point toward the same underlying physical process being at work. In this context, a comparison could also be made on the properties of X-ray ``shots'' (or quasi-flares, a wording used by \citealt{smi16}). Such ``shots'' were found to be not only ubiquitous but also gray (i.e. spectral properties were similar whatever the recorded amplitude) for the two \gc\ cases with sufficient signal-to-noise to study them in detail; that is, \gc\ itself and HD\,110432 \citep{smi98,smi12}. In contrast, only a few accreting WDs are found to display a strong flaring behavior. It is interpreted as being due to highly inhomogeneous (``blobby'') accretion and it usually occurs during bright and soft X-ray-dominated states (e.g. the soft polars V1309\,Ori, \citealt{sch05}, AI\,Tri, \citealt{tra10}, and QS\,Tel, \citealt{tra11}). A last case displaying ``shots'' is AE\,Aqr, but it corresponds to a propeller stage, not a typical accreting WD case \citep{ito06}. The light curves of these systems exhibit very different color temperatures during flares and in the background flux (\citealt{bea97,sch05} - see also discussion in \citealt{smi16}).

It is also usual to find in the light curves of CVs a modulation with either the WD spin or the orbital period, but no such variations have been detected for \gc\ analogs up to now \citep{smi16}. Regarding the spin signal, however, it remains to be examined what the expected WD spin would be in the case of \gc\ analogs as the known \gc\ binaries display much longer orbital periods than CVs. In fact, this will sensitively depend on the accretion parameters. The starting point may be a slow spin. Indeed, isolated WDs have a mean rotation period of 35\,hr with a standard deviation of 28\,hr \citep{hermes2017}, and the mass transfer from the WD progenitor onto the (future) Be star may have led to a further spin-down of the WD remnant. However, if accretion onto the WD occurs, this will rapidly spin it up, which is why we now examine this process. This requires one to consider whether an accretion disk is formed. To the first order, the conditions to form such an accretion disk do not depend on the nature of the compact object. Several 3D smooth particle hydrodynamics (SPH) models have studied such conditions for compact companions in the framework of Be/X-ray binaries. In most cases an accretion disk can indeed be formed around the neutron star \citep{oka2013,haya2004}, although \citet{martin2014} do not find evidence of a disk in their simulations. Observational evidence such as the detection of a soft excess probably due to the X-ray heated disk remains ambiguous \citep{hic2004}. However, the spin-up witnessed during outburst and spin-down in quiescence are best explained by the coupling between the disk and the magnetosphere \citep{gl1979}. To adapt to the case of WDs, which are larger than neutron stars, one needs to calculate the circularisation radius R$_{\rm circ}$. In order for an accretion disk to form, this radius must exceed the WD radius $R_{WD}$ and the effective size of its magnetosphere $R_{\rm M}$. \citet{frank2002} proposed as a condition for disk formation $R_{\rm circ} \geq 0.37\,R_{\rm M}$, with $R_{\rm circ} \geq R_{WD}$ and $R_{\rm circ}\,=\,J^{2}/GM_{\rm X}$ (where $J$ is the initial specific angular momentum of the accreted material). The magnetospheric radius is $R_{\rm M} = 5.5\,\times\,10^{8}\,M_{WD}\,R_{9}^{-2/7}\,L_{33}^{-2/7}\,\mu_{30}^{4/7}$\,cm \citep{frank2002}, where $R_{9}$ is the WD radius in units of 10$^{9}$\,cm, $L_{33}$ the X-ray luminosity in units of $10^{33}$\,erg\,s$^{-1}$ and $\mu_{30}$ the WD magnetic moment in units of $10^{30}$\,G cm$^{3}$. The angular momentum $J$ cannot be accurately calculated without resorting to hydrodynamical modeling. The SPH modelings of 4U 0115+63 by \citet{haya2004} and of A\,0535+262 by \citet{oka2013} in coplanar configurations suggest $R_{\rm circ}\,\sim\,0.01 \times a$ where $a$ is the semi-major axis of the orbit. If similar values of $R_{\rm circ}$ apply to \zt, for which $a\sim 252$\,R$_{\odot}$, only WDs with very high magnetic fields (like those seen in polars, see e.g. \citealt{fer2015}) could prevent the formation of an accretion disk. One can thus reasonably expect spin-up because of this disk. Assuming that $\Omega_{\rm WD}\,\ll\,\Omega_{\rm MAX}$ where $\Omega_{\rm MAX}$ is the breakup angular velocity, the angular velocity increase is given by $\dot{\Omega} = I^{-1} \dot{M} (G M_{WD} R)^{1/2}$, with $I$ the WD moment of inertia and $R$ the WD radius if accretion occurs through a disk extending down to the surface or the magnetospheric radius if the accretion disk is truncated by the WD magnetic field. In the slow rotator case, the spin-up rate can be written as
\begin{equation} \nonumber
\dot{f}_{disk}\,\simeq\,3.64 \times 10^{-18}\,I_{50}\,\dot{M}_{10}\,(R_{9}\,M_{WD})^{1/2} \, \rm{Hz\,s^{-1}} \\
\end{equation}
for disk accretion and 
\begin{equation} \nonumber
\dot{f}_{mag}\,\simeq\,1.96 \times 10^{-18}\,I_{50}\,\dot{M}_{10}^{6/7}\,M_{WD}^{3/7}\,\mu_{30}^{2/7} \, \rm{Hz\,s^{-1}}
\end{equation}
for magnetic accretion \citep{frank2002}. $I_{50}$ is the WD moment of inertia in units of $10^{50}$\,g\,cm$^{-2}$ ($I_{50} \approx 1.0$), and $\dot{M}_{10}$ the mass accretion rate in units of $10^{-10}$\,M$_{\odot}$\,yr$^{-1}$. Starting from the extreme case of a non-rotating WD, disk accretion yields $P_{spin}(s)\,\simeq\,8700\,(t/1Myr)^{-1}\,\dot{M}_{10}^{-1}$ while $P_{spin}(s)\,\simeq\,2200\,(t/1Myr)^{-1}\,\dot{M}_{10}^{-6/7}$ for accretion linked to a magnetically truncated disk with $B \sim 10^{6}$\,G, typical of IPs \citep{fer2015}. Therefore, for mass accretion rates of the order of $10^{-9}$\,M$_{\odot}$\,yr$^{-1}$, typical of high-luminosity IPs \citep{demartino2020}, X-ray pulsations with periods similar to those of IPs in equilibrium ($P_{spin}\sim 300-3000$\,s, \citealt{demartino2020}) should already be detectable for \zt\ (and more generally for \gc\ analogs) after only one Myr of activity. As no such stable X-ray period attributable to WD spins has been observed in \gc\ analogs, this configuration can probably be discarded. In the case of disk accretion, high accretion rates (hence high spin-up rates) are also possible: observations of dwarf novae in outburst or nova-like objects could imply accretion rates of up to a few $10^{-8}$\,M$_{\odot}$\,yr$^{-1}$ due to the low hard X-ray efficiency of the boundary layer \citep{bal20}. However, the WD rotation usually is undetectable for such systems in X-rays \citep{bal20} so that this configuration cannot be a priori excluded for \gc\ systems. 

A last parameter needs to be examined: absorption. In the context of X-rays born close to the WD companion of the Be star, there are two possible sources of absorption. One is due to the decretion disk of the Be star that crosses the line of sight toward the companion in some phases - and only some phases, since no such absorption should occur when the companion appears in front of the Be star. The absence of phase-locked variations in the absorbing column of \zt\ was noted earlier, but another point can be made in this context. Using the same disk model as in Section 4.2, the absorbing column toward an emission region located in the equatorial plane, at 252\,R$_{\odot}$ of the Be star (i.e. at the expected distance of the companion in \zt, \citealt{rud09,rau24}), can be calculated. For the favored inclination angle of 85$^{\circ}$ and considering the worst case (i.e. with the Be star in front of its companion), we find values of $0.6-8 \times 10^{22}$\,cm$^{-2}$, depending on the chosen set of disk parameters. Lower inclinations (75--80$^{\circ}$) yield much lower columns ($<2 \times 10^{22}$\,cm$^{-2}$), while inclinations close to 90$^{\circ}$ would lead to an eclipse of the X-ray emission zone, which has not been detected. The predicted absorption due to the disk should thus be (much) lower than observed which, combined to the absence of phase-locked modulation, suggests that the accretion scenario requires another absorber, close to the WD, to ensure orbital phase independence. The best candidate is of course linked to the accreting flows themselves and we now examine observational results regarding such absorption. 

In non-magnetic CVs such as dwarf novae, the disk extends down to the WD surface and X-rays are emitted from a boundary layer that may be optically thin at low to medium accretion rates with temperatures similar to those of \gc\ analogs \citep{patterson1985}. Since we assumed decretion and accretion disks to be virtually coplanar and since \zt\ is seen under a high inclination, we focus on cases of high-inclination CVs. A select number of dwarf novae in the quiescent state exhibit high inclination \citep{muk17}. For instance, \xmm\ observations of OY\,Car ($i = 83^{\circ}$, $L_{\rm X} \sim 5 \times 10^{30}$\,erg\,s$^{-1}$) require $N_{\rm H} \sim 10^{22}$\,cm$^{-2}$, with a covering fraction of $\sim50$\% \citep{pandel2005}. HT\,Cas ($i = 81^{\circ}$, $L_{\rm X} = 1.33 \times 10^{31}$\,erg\,s$^{-1}$) displays $N_{\rm H}\sim 1.6 - 3.3 \times 10^{21}$\,cm$^{-2}$ \citep{nucita2009}. In a similar manner, the X-ray spectrum of the partially eclipsing dwarf nova V893\,Sco ($i \sim 74^{\circ}$, $L_{\rm X} \sim 1.1 \times 10^{32}$\,erg\,s$^{-1}$) shows $N_{\rm H}\sim 3-6 \times 10^{21}$\,cm$^{-2}$ with a covering fraction of $\sim55$\% \citep{mukai2009}. These observations suggest that, among dwarf novae, the optically thin X-ray emitting boundary layer extends above the optically thick accretion disk which therefore does not block the view toward the emitting region \citep{muk17}: any emission arising there is not fully absorbed by the disk. Noticeably, the hard X-ray luminosities radiated by such sources in quiescence often are one or two orders of magnitude below those emitted by \gc\ analogs. For larger mass accretion rates, the boundary layer becomes optically thick, leading to enhanced UV and soft X-ray emission. With their $\sim100$-fold higher $\dot{M}$, dwarf novae in outburst and nova-like objects still exhibit hard X-ray luminosities in the range of $10^{29}$ to $10^{32}$\,erg\,s$^{-1}$ (see e.g. \citealt{bal20} and \citealt{sul2022}). In the few cases documented, nova-like X-ray spectra require $N_{\rm H} \leq 10^{22}$\,cm$^{-2}$ \citep{zemko2014}. It is therefore unclear how disk accretion onto a WD could account for the high $N_{\rm H}$, high temperature, and high intrinsic X-ray luminosity of \zt.

If the WD magnetic field strength is high enough to drive accretion onto its magnetic poles, the inner part of the accretion disk may be truncated as in IPs. For even higher magnetic fields (as in polars), matter falls ballistically from L1 onto the WD and is later captured by the magnetosphere without forming an accretion disk (see \citealt{demartino2020} for a recent review). In such systems, most of the photoelectric absorption takes place in the pre-shock material located above the X-ray emitting region of the accretion column. Since X-rays emitted in the shock cross different amounts of pre-shock matter, the emerging X-ray spectrum is best described by a partially covering absorber or a more physically motivated distribution of absorbing columns \citep{done1998}. The \xmm\ survey of IPs carried out by \citet{bernardini2017} reveals the presence of thick photoelectric absorbers with $N_{\rm H}$ in the range of $1.6 - 8.7 \times 10^{23}$\,cm$^{-2}$ with covering fractions spanning from 33 to 76\%. Additional orbital phase-dependent absorption often occurs in IPs at a level of $10^{22}$\,cm$^{-2}$ \citep{parker2005}. This absorbing component is believed to be due to the bulge created by the stream of matter impacting the external part of the accretion disk. Few IPs are known to display eclipses testifying to a high inclination. The best documented case is XY\,Ari \citep{zen2018}. Best fits to its X-ray spectrum require a complex absorption pattern consisting of a global $N_{\rm H}$ of $\sim 3 \times 10^{22}$\,cm$^{-2}$ and two partial absorbers with $N_{\rm H}  \sim 10^{24}$\,cm$^{-2}$ and $\sim 6 \times 10^{22}$\,cm$^{-2}$ with covering fractions of 41 and 53\% respectively. Polars exhibit similar partial absorptions with coverage of $\sim$50\% and $N_{\rm H}$ of up to a few $10^{23}$\,cm$^{-2}$ \citep{bernardini2014,tra14,rawat2023}. Additional narrow absorption features due to streams of matter crossing the line of sight may also occur as in AM\,Her \citep{schwope2020} with a further absorbing column of $4.1 \times 10^{23}$\,cm$^{-2}$ and a covering fraction of 93\%. EX\,Hydrae is an intermediate case between IPs and polars where the WD has a 67\,min spin period not yet synchronized with the 98\,min orbital period. It also shows partial eclipses by the mass donor star. In spite of the high inclination, photoelectric absorption reaches only a few $10^{21}$\,cm$^{-2}$ \citep{allan1998}. In fact, since the bulk of the photoelectric absorption in magnetic CVs occurs in the stream of matter driven by the strong magnetic field, there should be no inclination effect as the magnetic field orientation is a priori unrelated to that of the accretion disk. Indeed, high inclination systems do not exhibit much enhanced absorption, and absorption is always partial. This seems at odds with the observations of \gc\ in general and \zt\ in particular, with its higher absorbing column. This is a second argument, after the non-detection of stable WD spin signals, against the magnetic configuration.

A small group of symbiotic stars, called ``$\delta$-type'' objects,  emit optically thin thermal hard X-rays with temperatures in the range of 15 to 30\,keV and X-ray luminosities of $\sim 10^{33}$\,erg\,s$^{-1}$ \citep{ken09}, comparable with those of \gc\ analogs. Global intrinsic absorption with $N_{\rm H}\leq 2 \times 10^{22}$\,cm$^{-2}$ is observed, in addition to partially covering components of up to $\sim 3 \times 10^{23}$\,cm$^{-2}$ with covering fractions as high as 96\% \citep{ken09}. Although accretion mostly occurs through stellar wind, a disk may form around the WD \citep{bal20}. For example, the hard component of T\,CrB has an X-ray luminosity of $\sim 6\times 10^{32}$\,erg\,s$^{-1}$ and is best fitted with a cooling flow ($kT_{\rm Max}\approx 15$\,keV) absorbed by $N_{\rm H} = 3-7 \times 10^{23}$\,cm$^{-2}$ with an extremely high partial covering factor of 99.7\% \citep{luna2018,zhe2019}. A luminous soft black body appears in the active state, and so these authors interpret the soft and hard X-ray component as being due to the emission of the boundary layer, with the change of state of T\,CrB being equivalent to the quiescent to outburst transition in dwarf novae. However, RT\,Cru, another $\delta$-type symbiotic star, fails to show the dwarf nova behavior of T\,CrB \citep{luna2018b}. Classical CVs do not exhibit such a high covering factor of their hard X-ray component, suggesting that the accretion geometry in $\delta$-type symbiotic stars is significantly different, in particular the visibility of the boundary layer. It is striking that the observed absorption and covering factors of T\,CrB are almost identical with those we derive for \zt . However, if the symbiotic star scenario applies to \gc\ analogs, the rather moderate ($\sim 60^{\circ}$) inclination of T\,CrB \citep{bel1998} would imply similarly high absorptions and covering factors in other \gc\ analogs (which are seen at lower inclinations than \zt). This is not observed. Moreover, in the absence of known eclipsing $\delta$-type symbiotic stars, the role played by the accretion disk at high inclinations remains uncertain. While symbiotics are the most promising WD proxies for \gc\ stars, much remains to be done to clarify their behavior in \gc-like orbital geometries.

\section{Conclusions and summary}
It was recently discovered that the \zt\ binary displays the peculiar X-ray characteristics of \gc\ analogs. To further characterize its properties, the system was the target of an \xmm\ campaign. The exposures were obtained at two specific orbital phases, first when the Be star (whose disk is seen edge-on) was in front of its companion and then in the opposite orbital configuration. These X-ray observations were complemented by archival X-ray (\ch, eROSITA) and optical (\te, BeSS) data.

The \xmm\ light curves present significant changes occurring with various timescales. As with other \gc\ stars, no persistent periodic signal is detected. Rather, the variability amplitude appears to decrease with increasing frequency, the power spectrum approximately following $P\propto 1/f$ as was found in \gc. The ratio between count rates in the 1.5--4.0 and 4.0--10.0\,keV energy bands varies from epoch to epoch, but it remains stable during each of the \xmm\ exposures: the variations are thus gray on short timescales. The ratio between count rates in 0.5--1.5 and 4.0--10.0\,keV is noisier and seems to slightly decrease at the end of the first \xmm-pn exposure. This may suggest the occurrence of a transient, shallow ``softness dip'' that could be related to a slight increase in absorption toward the warm plasma. In the optical, \te\ data of \zt\ seem quite typical for a Be star, with broad frequency groups appearing at low frequencies and no significant high-frequency (10--360\,d$^{-1}$) signal. 

The X-ray spectrum presents a very faint soft component and a bright hard component. The intensity of the former appears in line with those recorded for non-\gc\ Be stars and could therefore correspond to intrinsic X-rays from the wind of the massive star. The hard component suffers from a strong absorbing column (at least $10^{23}$\,cm$^{-2}$) that significantly varies up to a factor of a few. This absorption variation does not appear phase-locked, nor is it correlated with the properties (overall strength, (a)symmetry, depth of the core absorption) of the H$\alpha$ line, the usual Be disk tracer.

These observed properties challenge the proposed scenarios to explain the X-ray emission of \gc\ analogs. The observed absence of orbital modulation is expected for a star-disk interaction scenario. In addition, absorptions similar to the observed values can be reached in that scenario if the X-ray emission arises from near the star although not exactly at its equator but slightly above it. However, it remains to be seen how to reconcile this scenario with no (or little) correlation between X-ray and disk properties. For the accreting WD scenario, one could draw similarities between \gc\ analogs and some known accreting WDs regarding for example the frequency behavior of the stochastic variability, the high plasma temperature, the presence of a fluorescence line, and even the strong absorption. However, it remains unclear if the full set of characteristics observed in \gc\ analogs can be found altogether in a single type of accreting WD. For example, CVs may display strong absorption, but only with a partial coverage, contrary to what is detected for \zt. In addition, stable short-term modulations typically expected to be linked to the WD spin in IPs have not been detected in \gc\ analogs, nor any orbital modulation or any outburst typical of novae. It must be noted, however, that the detailed modeling of both scenarios for \gc-like configurations is still lacking, prohibiting a thorough testing. The availability of such modeling clearly constitutes an essential step in our way toward understanding \gc\ analogs.

\begin{acknowledgements}
Y.N. and G.R. acknowledge support from the Fonds National de la Recherche Scientifique (Belgium), the European Space Agency (ESA) and the Belgian Federal Science Policy Office (BELSPO) in the framework of the PRODEX Programme (contracts linked to XMM-Newton). M.A.S. acknowledges support from \ch\ grant \#362675 to Catholic University of America. J.R. acknowledges support from the DLR under grant 50QR2105. ADS and CDS were used for preparing this document. This work has made use of the AAVSO repository (https://www.aavso.org) and BeSS database, operated at LESIA, Observatoire de Meudon, France (http://basebe.obspm.fr). This work uses data from eROSITA, the soft X-ray instrument aboard SRG.
\end{acknowledgements}


\begin{thebibliography}{}
\bibitem[Asplund et al.(2009)]{asp09} Asplund, M., Grevesse, N., Sauval, A.~J., et al.\ 2009, \araa, 47, 481. 
\bibitem[Allan et al.(1998)]{allan1998} Allan, A., Hellier, C., \& Beardmore, A.\ 1998, \mnras, 295, 167. 
\bibitem[Balman(2019)]{bal19} Balman, {\c{S}}.\ 2019, Astronomische Nachrichten, 340, 296. 
\bibitem[Balman(2020)]{bal20} Balman, {\c{S}}.\ 2020, Advances in Space Research, 66, 1097. 
\bibitem[Beardmore \& Osborne(1997)]{bea97} Beardmore, A.~P. \& Osborne, J.~P.\ 1997, \mnras, 290, 145.
\bibitem[Belczynski \& Mikolajewska(1998)]{bel1998} Belczynski, K. \& Mikolajewska, J.\ 1998, \mnras, 296, 77. 
\bibitem[Bernardini et al.(2014)]{bernardini2014} Bernardini, F., de Martino, D., Mukai, K., et al.\ 2014, \mnras, 445, 1403. 
\bibitem[Bernardini et al.(2017)]{bernardini2017} Bernardini, F., de Martino, D., Mukai, K., et al.\ 2017, \mnras, 470, 4815.
\bibitem[Carciofi et al.(2009)]{car09} Carciofi, A.~C., Okazaki, A.~T., Le Bouquin, J.-B., et al.\ 2009, \aap, 504, 915. 
\bibitem[Christian(2000)]{chr00} Christian, D.~J.\ 2000, \aj, 119, 1930. 
\bibitem[Cochetti et al.(2019)]{coc19} Cochetti, Y.~R., Arcos, C., Kanaan, S., et al.\ 2019, \aap, 621, A123. 
\bibitem[Coe et al.(2020)]{coe20} Coe, M.~J., Kennea, J.~A., Evans, P.~A., et al.\ 2020, \mnras, 497, L50. 
\bibitem[Cracco et al.(2018)]{cra18} Cracco, V., Orio, M., Ciroi, S., et al.\ 2018, \apj, 862, 167. 
\bibitem[de Martino et al.(2020)]{demartino2020} de Martino, D., Bernardini, F., Mukai, K., et al.\ 2020, Advances in Space Research, 66, 1209.
\bibitem[Done \& Magdziarz(1998)]{done1998} Done, C. \& Magdziarz, P.\ 1998, \mnras, 298, 737. 
\bibitem[Ferrario et al.(2015)]{fer2015} Ferrario, L., de Martino, D., \& G{\"a}nsicke, B.~T.\ 2015, \ssr, 191, 111. 
\bibitem[Frank et al.(2002)]{frank2002} Frank, J., King, A., \& Raine, D.~J.\ 2002, Accretion Power in Astrophysics, by Juhan Frank and Andrew King and Derek Raine, pp. 398. ISBN 0521620538. Cambridge, UK: Cambridge University Press, February 2002., 398
\bibitem[Gies et al.(2007)]{gie07} Gies, D.~R., Bagnuolo, W.~G., Baines, E.~K., et al.\ 2007, \apj, 654, 527. 
\bibitem[Gies et al.(2023)]{gie23} Gies, D.~R., Wang, L., \& Klement, R.\ 2023, \apjl, 942, L6. 
\bibitem[Ghosh \& Lamb(1979)]{gl1979} Ghosh, P. \& Lamb, F.~K.\ 1979, \apj, 234, 296. 
\bibitem[Gosset et al.(2001)]{gos01} Gosset, E., Royer, P., Rauw, G., et al.\ 2001, \mnras, 327, 435. 
\bibitem[Grundstrom \& Gies(2006)]{gru06} Grundstrom, E.~D. \& Gies, D.~R.\ 2006, \apjl, 651, L53. 
\bibitem[Hamaguchi et al.(2016)]{ham16} Hamaguchi, K., Oskinova, L., Russell, C.~M.~P., et al.\ 2016, \apj, 832, 140. 
\bibitem[Hayasaki et al.(2004)]{haya2004} Hayasaki, K. \& Okazaki, A.~T.\ 2004, \mnras, 350, 971. 
\bibitem[Hermes et al.(2017)]{hermes2017} Hermes, J.~J., G{\"a}nsicke, B.~T., Kawaler, S.~D., et al.\ 2017, \apjs, 232, 23. 
\bibitem[Hickox et al.(2004)]{hic2004} Hickox, R.~C., Narayan, R., \& Kallman, T.~R.\ 2004, \apj, 614, 881. 
\bibitem[Itoh et al.(2006)]{ito06} Itoh, K., Okada, S., Ishida, M., et al.\ 2006, \apj, 639, 397. 
\bibitem[Kahabka et al.(2006)]{kah06} Kahabka, P., Haberl, F., Payne, J.~L., et al.\ 2006, \aap, 458, 285. 
\bibitem[Kennea et al.(2009)]{ken09} Kennea, J.~A., Mukai, K., Sokoloski, J.~L., et al.\ 2009, \apj, 701, 1992. 
\bibitem[Kennea et al.(2021)]{ken21} Kennea, J.~A., Coe, M.~J., Evans, P.~A., et al.\ 2021, \mnras, 508, 781. 
\bibitem[Labadie-Bartz et al.(2022)]{lab22} Labadie-Bartz, J., Carciofi, A.~C., Henrique de Amorim, T., et al.\ 2022, \aj, 163, 226. 
\bibitem[Langer et al.(2020)]{lan20} Langer, N., Baade, D., Bodensteiner, J., et al.\ 2020, \aap, 633, A40. 
\bibitem[Li et al.(2012)]{li12} Li, K.~L., Kong, A.~K.~H., Charles, P.~A., et al.\ 2012, \apj, 761, 99. 
\bibitem[Lopes de Oliveira et al.(2010)]{lop10} Lopes de Oliveira, R., Smith, M.~A., \& Motch, C.\ 2010, \aap, 512, A22. 
\bibitem[Luna et al.(2018)]{luna2018} Luna, G.~J.~M., Mukai, K., Sokoloski, J.~L., et al.\ 2018, \aap, 619, A61. 
\bibitem[Luna et al.(2018)]{luna2018b} Luna, G.~J.~M., Mukai, K., Sokoloski, J.~L., et al.\ 2018, \aap, 616, A53. 
\bibitem[Martin et al.(2014)]{martin2014} Martin, R.~G., Nixon, C., Armitage, P.~J., et al.\ 2014, \apjl, 790, L34. 
\bibitem[Motch et al.(2015)]{mot15} Motch, C., Lopes de Oliveira, R., \& Smith, M.~A.\ 2015, \apj, 806, 177. 
\bibitem[Mukai et al.(2009)]{mukai2009} Mukai, K., Zietsman, E., \& Still, M.\ 2009, \apj, 707, 652. 
\bibitem[Mukai(2017)]{muk17} Mukai, K.\ 2017, \pasp, 129, 062001. 
\bibitem[Naz{\'e} et al.(2017)]{naz17} Naz{\'e}, Y., Rauw, G., \& Cazorla, C.\ 2017, \aap, 602, L5. 
\bibitem[Naz{\'e} \& Motch(2018)]{naz18} Naz{\'e}, Y. \& Motch, C.\ 2018, \aap, 619, A148. 
\bibitem[Naz{\'e} et al.(2019)]{naz19} Naz{\'e}, Y., Rauw, G., \& Smith, M.\ 2019, \aap, 632, A23. 
\bibitem[Naz{\'e} et al.(2020)]{naz20} Naz{\'e}, Y., Rauw, G., \& Pigulski, A.\ 2020, \mnras, 498, 3171. 
\bibitem[Naz{\'e} et al.(2022a)]{naz22bin} Naz{\'e}, Y., Rauw, G., Czesla, S., et al.\ 2022a, \mnras, 510, 2286. 
\bibitem[Naz{\'e} et al.(2022b)]{naz22mon} Naz{\'e}, Y., Rauw, G., Bohlsen, T., et al.\ 2022b, \mnras, 512, 1648. 
\bibitem[Naz{\'e} et al.(2022c)]{naz22} Naz{\'e}, Y., Rauw, G., Smith, M.~A., et al.\ 2022c, \mnras, 516, 3366. 
\bibitem[Naz{\'e} \& Robrade(2023)]{naz23} Naz{\'e}, Y. \& Robrade, J.\ 2023, \mnras, 525, 4186. 
\bibitem[Naz{\'e} et al.(2024)]{naz24} Naz{\'e}, Y., et al.\ 2024, OEJV, 246
\bibitem[Nucita et al.(2009)]{nucita2009} Nucita, A.~A., Maiolo, B.~M.~T., Carpano, S., et al.\ 2009, \aap, 504, 973. 
\bibitem[Okazaki et al.(2013)]{oka2013} Okazaki, A.~T., Hayasaki, K., \& Moritani, Y.\ 2013, \pasj, 65, 41. 
\bibitem[Orio et al.(2010)]{ori10} Orio, M., Nelson, T., Bianchini, A., et al.\ 2010, \apj, 717, 739. 
\bibitem[Pandel et al.(2005)]{pandel2005} Pandel, D., C{\'o}rdova, F.~A., Mason, K.~O., et al.\ 2005, \apj, 626, 396. 
\bibitem[Parker et al.(2005)]{parker2005} Parker, T.~L., Norton, A.~J., \& Mukai, K.\ 2005, \aap, 439, 213. 
\bibitem[Patterson \& Raymond(1985)]{patterson1985} Patterson, J. \& Raymond, J.~C.\ 1985, \apj, 292, 535. 
\bibitem[Pollmann(2017)]{pol17} Pollmann, E.\ 2017, Information Bulletin on Variable Stars, 6208, 1. 
\bibitem[Pols et al.(1991)]{pol91} Pols, O.~R., Cote, J., Waters, L.~B.~F.~M., et al.\ 1991, \aap, 241, 419
\bibitem[Postnov et al.(2017)]{pos17} Postnov, K., Oskinova, L., \& Torrej{\'o}n, J.~M.\ 2017, \mnras, 465, L119. 
\bibitem[Quirrenbach et al.(1997)]{qui97} Quirrenbach, A., Bjorkman, K.~S., Bjorkman, J.~E., et al.\ 1997, \apj, 479, 477. 
\bibitem[Rauw et al.(2022)]{rau22} Rauw, G., Naz{\'e}, Y., Motch, C., et al.\ 2022, \aap, 664, A184. 
\bibitem[Rauw (2024)]{rau24} Rauw, G.\ 2024, \aap, 682, A179
\bibitem[Rawat et al.(2023)]{rawat2023} Rawat, N., Pandey, J.~C., Joshi, A., et al.\ 2023, \mnras, 521, 2729. 
\bibitem[Revnivtsev et al.(2011)]{rev11} Revnivtsev, M., Potter, S., Kniazev, A., et al.\ 2011, \mnras, 411, 1317. 
\bibitem[Ru{\v{z}}djak et al.(2009)]{rud09} Ru{\v{z}}djak, D., Bo{\v{z}}i{\'c}, H., Harmanec, P., et al.\ 2009, \aap, 506, 1319. 
\bibitem[Schaefer et al.(2010)]{sch10} Schaefer, G.~H., Gies, D.~R., Monnier, J.~D., et al.\ 2010, \aj, 140, 1838. 
\bibitem[Schwarz et al.(2005)]{sch05} Schwarz, R., Reinsch, K., Beuermann, K., et al.\ 2005, \aap, 442, 271. 
\bibitem[Schwope et al.(2020)]{schwope2020} Schwope, A.~D., Worpel, H., Traulsen, I., et al.\ 2020, \aap, 642, A134. 
\bibitem[Shao \& Li(2014)]{sha14} Shao, Y. \& Li, X.-D.\ 2014, \apj, 796, 37. 
\bibitem[Smith et al.(1998)]{smi98} Smith, M.~A., Robinson, R.~D., \& Corbet, R.~H.~D.\ 1998, \apj, 503, 877.  
\bibitem[Smith et al.(2004)]{smi04} Smith, M.~A., Cohen, D.~H., Gu, M.~F., et al.\ 2004, \apj, 600, 972. 
\bibitem[Smith et al.(2012a)]{smi12} Smith, M.~A., Lopes de Oliveira, R., \& Motch, C.\ 2012a, \apj, 755, 64. 
\bibitem[Smith et al.(2012b)]{smi12b} Smith, M.~A., Lopes de Oliveira, R., Motch, C., et al.\ 2012b, \aap, 540, A53. 
\bibitem[Smith et al.(2016)]{smi16} Smith, M.~A., Lopes de Oliveira, R., \& Motch, C.\ 2016, Advances in Space Research, 58, 782
\bibitem[Smith et al.(2017)]{smi17} Smith, M.~A., Lopes de Oliveira, R., \& Motch, C.\ 2017, \mnras, 469, 1502. 
\bibitem[Smith \& Lopes de Oliveira(2019)]{smi19} Smith, M.~A. \& Lopes de Oliveira, R.\ 2019, \mnras, 488, 5048. 
\bibitem[{\v{S}}tefl et al.(2009)]{ste09} {\v{S}}tefl, S., Rivinius, T., Carciofi, A.~C., et al.\ 2009, \aap, 504, 929. 
\bibitem[Sturm et al.(2012)]{stu12} Sturm, R., Haberl, F., Pietsch, W., et al.\ 2012, \aap, 537, A76.
\bibitem[Suleimanov et al.(2022)]{sul2022} Suleimanov, V.~F., Doroshenko, V., \& Werner, K.\ 2022, \mnras, 511, 4937. 
\bibitem[van Bever \& Vanbeveren(1997)]{van97} van Bever, J. \& Vanbeveren, D.\ 1997, \aap, 322, 116
\bibitem[Touhami et al.(2011)]{tou11} Touhami, Y., Gies, D.~R., \& Schaefer, G.~H.\ 2011, \apj, 729, 17. 
\bibitem[Touhami et al.(2013)]{tou13} Touhami, Y., Gies, D.~R., Schaefer, G.~H., et al.\ 2013, \apj, 768, 128. 
\bibitem[Torrej{\'o}n et al.(2012)]{tor12} Torrej{\'o}n, J.~M., Schulz, N.~S., \& Nowak, M.~A.\ 2012, \apj, 750, 75. 
\bibitem[Traulsen et al.(2010)]{tra10} Traulsen, I., Reinsch, K., Schwarz, R., et al.\ 2010, \aap, 516, A76. 
\bibitem[Traulsen et al.(2011)]{tra11} Traulsen, I., Reinsch, K., Schwope, A.~D., et al.\ 2011, \aap, 529, A116. 
\bibitem[Traulsen et al.(2014)]{tra14} Traulsen, I., Reinsch, K., Schwope, A.~D., et al.\ 2014, \aap, 562, A42. 
\bibitem[Tycner et al.(2004)]{tyc04} Tycner, C., Hajian, A.~R., Armstrong, J.~T., et al.\ 2004, \aj, 127, 1194. 
\bibitem[Wang et al.(2021)]{wan21} Wang, L., Gies, D.~R., Peters, G.~J., et al.\ 2021, \aj, 161, 248. 
\bibitem[Xu et al.(2016)]{xu16} Xu, X.-j., Wang, Q.~D., \& Li, X.-D.\ 2016, \apj, 818, 136. 
\bibitem[Zemko et al.(2014)]{zemko2014} Zemko, P., Orio, M., Mukai, K., et al.\ 2014, \mnras, 445, 869. 
\bibitem[Zengin {\'C}amurdan et al.(2018)]{zen2018} Zengin {\'C}amurdan, D., Balman, {\c{S}}., \& Burwitz, V.\ 2018, \mnras, 477, 4035. 
\bibitem[Zhekov \& Tomov(2019)]{zhe2019} Zhekov, S.~A. \& Tomov, T.~V.\ 2019, \mnras, 489, 2930. 
\end{thebibliography}
\end{document}